\documentclass[11pt,preprint]{aastex}

\newcommand{\Te} {T_{\rm eff}}
\newcommand{\logg} {\log g}

\begin{document}

\title{WHITE DWARFS IN THE UKIRT INFRARED DEEP SKY SURVEY DATA RELEASE 9}

\author{P.-E. Tremblay$^{1,2}$, S.~K. Leggett$^{3}$,
  N. Lodieu$^{4,5}$, B. Freytag$^{6}$, P. Bergeron$^{7}$,
  J.~S. Kalirai$^{1}$, and H.-G. Ludwig$^{8}$}

\affil{$^{1}$Space Telescope Science Institute, 700 San Martin Drive,
  Baltimore, MD, 21218, USA} \affil{$^{2}$Hubble Fellow}
\email{tremblay@stsci.edu} \affil{$^{3}$Gemini Observatory, Northern
  Operations Center, 670 North A'ohoku Place, Hilo, HI, 96720, USA}
\affil{$^{4}$Instituto de Astrof\'isica de Canarias (IAC), C/ V\'ia
  L\'actea s/n, E-38200 La Laguna, Tenerife, Spain}
\affil{$^{5}$Departamento de Astrof\'isica, Universidad de La Laguna
  (ULL), E-38205 La Laguna, Tenerife, Spain} \affil{$^{6}$Astronomical
  Observatory, Uppsala University, Regementsv\"agen 1, Box 515,
  SE-75120 Uppsala, Sweden} \affil{$^{7}$D\'epartement de Physique,
  Universit\'e de Montr\'eal, C. P. 6128, Succursale Centre-Ville,
  Montr\'eal, QC H3C 3J7, Canada} \and \affil{$^{8}$Zentrum f\"ur
  Astronomie der Universit\"at Heidelberg, Landessternwarte,
  K\"onigstuhl 12, Heidelberg, 69117, Germany}

\begin{abstract}

We have identified eight to ten new cool white dwarfs from the Large
Area Survey (LAS) Data Release 9 of the United Kingdom InfraRed
Telescope (UKIRT) Infrared Deep Sky Survey (UKIDSS). The data set was
paired with the Sloan Digital Sky Survey (SDSS) to obtain proper
motions and a broad $ugrizYJHK$ wavelength coverage. Optical
spectroscopic observations were secured at Gemini Observatory and
confirm the degenerate status for eight of our targets. The final
sample includes two additional white dwarf candidates with no
spectroscopic observations. We rely on improved 1D model atmospheres
and new multi-dimensional simulations with CO5BOLD to review the
stellar parameters of the published LAS white dwarf sample along with
our additional discoveries. Most of the new objects possess very cool
atmospheres with effective temperatures below 5000~K, including two
pure-hydrogen remnants with a cooling age between 8.5 and 9.0~Gyr, and
tangential velocities in the range 40~km~s$^{-1}$ $\leq v_{\rm tan}
\leq$ 60~km~s$^{-1}$. They are likely thick disk 10-11 Gyr-old
objects. In addition we find a resolved double degenerate system with
$v_{\rm tan} \sim$ 155 km~s$^{-1}$ and a cooling age between 3.0 and
5.0~Gyr. These white dwarfs could be disk remnants with a very high
velocity or former halo G stars. We also compare the LAS sample with
earlier studies of very cool degenerates and observe a similar deficit
of helium-dominated atmospheres in the range $5000 < T_{\rm eff}$ (K)
$< 6000$. We review the possible explanations for the spectral
evolution from helium-dominated towards hydrogen-rich atmospheres at
low temperatures.

\end{abstract}

\keywords{white dwarfs -- stars: fundamental properties -- infrared:
  stars -- techniques: photometric -- techniques: spectroscopic --
  surveys}

\section{INTRODUCTION}

The remnants of most intermediate mass stars (0.6 $ \lesssim
M/M_{\odot} \lesssim $ 8) are C/O core white dwarfs with a hydrogen or
helium-rich atmosphere. The cooling rate of white dwarfs decreases as
they age and the oldest remnants in the solar neighborhood have
atmospheres that are warmer than most stars in the same
environment. Due to their small radius, however, cool degenerate stars
are intrinsically faint and the local sample is only complete {\bf out} to
about 20 pc \citep{holberg08,sion09,giammichele12}.  Early studies of
the white dwarf local population have shown that the observed
luminosity function drops off for effective temperatures $T_{\rm eff}
\lesssim$ 4000~K, and corresponding ages in the range of 8-10 Gyr,
which is consistent with the formation time of the Galactic thin disk
\citep{winget87,liebert88,leggett98}.

Cool white dwarfs are unambiguously old since their time on the
cooling sequence is a significant fraction of the galactic age. Given
that the mass of a remnant is known, it is possible to derive the
initial mass of the star and the total age of the white dwarf by using
the initial-final mass relation calibrated from clusters \citep[see,
  e.g.,][]{kalirai08,williams09,dobbie12}. However, spectroscopic
observations of very cool white dwarfs, with $T_{\rm eff} \lesssim$
5000~K depending on the resolution, show no hydrogen or helium lines
(DC spectral type) that can be used to constrain the surface gravity.
Furthermore, photometric information is not sufficient to constrain
masses unless trigonometric parallax observations are available
\citep{bergeron97}. In order to identify thick disk or halo white
dwarfs in the solar neighborhood, recent studies have instead relied
on photometry coupled with kinematics
\citep{kilic09,kilic10,paper1,paper2}. They also secured spectroscopic
observations for most targets to confirm the degenerate
nature. Remnants that have halo or thick disk kinematics, along with
large cooling ages assuming a typical gravity of $\log g = 8$, are
then selected as likely candidates for those galactic
components. Follow-up parallax observations can confirm the membership
of these candidates by providing masses and, hence, total ages
\citep[see, e.g.,][]{kilic12}. It is hoped that kinematic,
photometric, and parallax data from the {\it Gaia} survey will
identify a significant halo and thick disk population {\bf of white
  dwarfs} \citep{carrasco13}.

In order to exploit the full potential of future surveys to study old
galactic populations, the accuracy of the cool white dwarf models must
be improved. One of the most pressing issues is to solve the
discrepancy between the observed and predicted collision-induced
absorption (CIA) features due to H$_2$ molecules. The current model
fits of cool degenerates with a derived pure-hydrogen atmosphere and
$T_{\rm eff} \lesssim 4000$~K, or a mixed helium-hydrogen (He/H)
atmosphere below $T_{\rm eff} \sim 6000$~K, are not fully satisfactory
\citep{bergeron02,kilic08,paper2,kilic12}, although it is hoped that
improved CIA opacity calculations \citep[see, e.g.,][]{frommhold10}
will solve this issue.

Another uncertain aspect is related to the chemical evolution of cool
white dwarfs. \citet{kowalski06} have found the previously missing
Lyman $\alpha$ (Ly$\alpha$) red wing opacity which has a strong impact
on the blue part of the visible flux (e.g. SDSS $u$ and $g$ bands) of
cool H-rich atmospheres ($T_{\rm eff} \lesssim 5500$~K). Recent
studies that have relied on this opacity \citep{kilic10,giammichele12}
suggest that most DCs below $T_{\rm eff} < 5000$~K appear to have
H-rich atmospheres. On the other hand, analyses without the Ly$\alpha$
opacity \citep{bergeron01,kilic09,paper2} derive a ratio of helium to
hydrogen atmospheres that is roughly 50\% in the same $T_{\rm eff}$
range, similar to what is observed at warmer temperatures \citep[$6000
  < T_{\rm eff}$ (K) $< 8000$;][]{tremblay08,giammichele12}. There is
currently no physical model able to explain the spectral evolution
towards H-rich atmospheres suggested by the photometric analyses
including the Ly$\alpha$ opacity. It is important to understand the
properties of this chemical evolution since the {\bf derived age
  estimates} of cool remnants can change by $\sim$1 Gyr depending on
the total amount of hydrogen present in the white dwarf
\citep{fontaine01}.

In this work, we use the United Kingdom InfraRed Telescope (UKIRT)
Infrared Deep Sky Survey \citep[UKIDSS;][]{lawrence07} Large Area
Survey (LAS) Data Release (DR)~9 coupled with the Sloan Digital Sky
Survey \citep[SDSS;][]{york00} to identify 18 white dwarf
  candidates, out of which eight are confirmed as degenerates from
  spectroscopy. We also identify two more possible white dwarfs
  without spectroscopic confirmation. We combine them with the white
dwarfs identified in previous LAS data releases \citep[][thereafter
  Paper~I and Paper II, respectively]{paper1,paper2} to form a sample
of 30 cool white dwarfs. We review the properties of this sample, in
particular the chemical abundances, by employing improved 1D model
atmospheres including the Ly$\alpha$ red-wing opacity. On the
theoretical side, we also present the first 3D model atmospheres of
very cool pure-H white dwarfs in order to estimate the 3D effects on
the predicted spectra. These hydrodynamical simulations may be the
first step in understanding the chemical evolution of cool white
dwarfs, which could be caused by events related to convection in the
deeper layers that are not currently part of our 3D simulations.

We begin in Section 2 by presenting an overview of the LAS sample of
cool white dwarfs from Paper~I and II as well as the white dwarf
candidates observed photometrically and spectroscopically in this
work. Section 3 describes our 1D and new 3D model atmospheres employed
to study these former stars. We continue in Section 4 by comparing the
objects in the LAS sample to model predictions in order to determine
updated stellar properties. We review the status of the spectral
evolution of cool white dwarfs in Section 5 and conclude in Section 6.

\section{SAMPLE SELECTION AND OBSERVATIONS}

The study presented here involves the white dwarfs identified from the
pairing of the UKIDSS LAS and SDSS surveys in Paper~I and II. We begin
with a brief review of the sample selection and observations in the
published sample, and then we present new candidate white dwarfs
identified from the LAS DR9.

\subsection{UKIDSS LAS Sample}

The LAS is a sub-survey of UKIDSS aimed at identifying cool and faint
galactic sources in the $YJHK$ bands \citep{hewett06} with a large
photometric depth. In Paper~I, target white dwarfs were selected in
the 280 deg$^2$ sky area of the LAS DR2, producing a sample of seven
new white dwarfs with $T_{\rm eff} \sim 6000$~K, confirmed with
optical spectroscopy obtained at Gemini Observatory. Paper~II used the
larger 1400 deg$^2$ area covered by the LAS DR6 and refined the color
selection to identify 13 more white dwarfs, including seven objects
with $T_{\rm eff} < 4500$~K.

Near-infrared photometry is essential to characterize cool white
dwarfs since the CIA due to H$_2$ molecules is observed in this
wavelength range. It is also crucial to have a broad wavelength
coverage in order to determine precise $T_{\rm eff}$ values. The LAS
and SDSS databases were therefore cross-correlated in order to add
$ugriz$ photometric information. Furthermore, proper motions were
derived using the target coordinates and the epochs of the SDSS and
UKIDSS images. Sources were matched by requiring the presence of a
primary SDSS source within 5$\arcsec$ of one LAS point source
coordinate.

\begin{figure}[p]
\epsscale{0.9}
\plotone{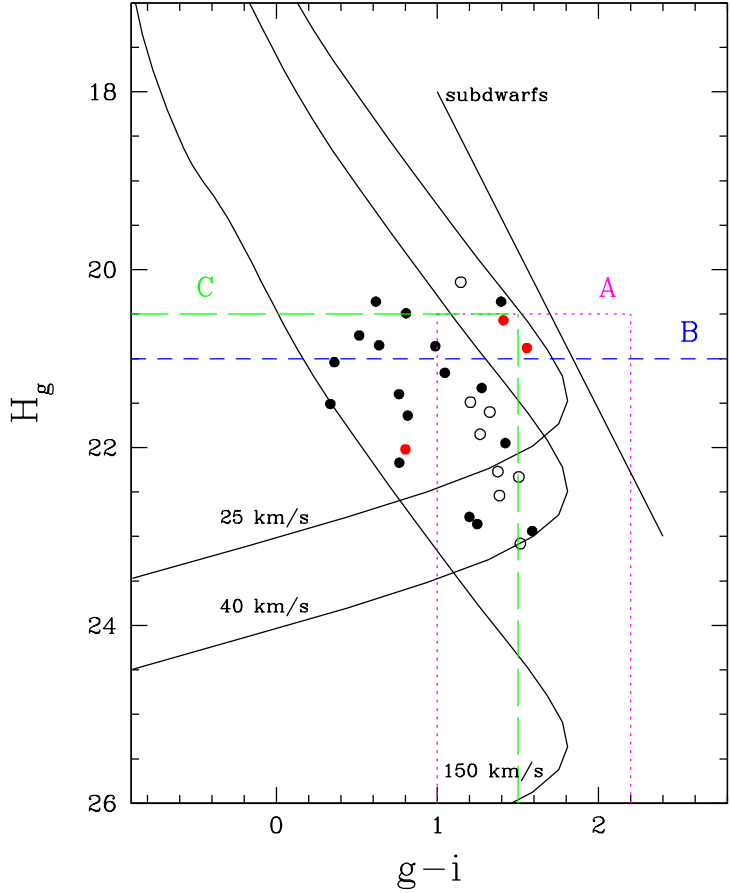}
\begin{flushright}
\caption{Reduced $g$-band proper motion as a function of $g-i$. We
  display the published sample of Paper~I and II (filled circles) and
  the white dwarfs discovered in this work (open circles). Mixed He/H
  objects are identified in red. White dwarf cooling curves (solid
  lines) for $v_{\rm tan}$ = 25 and 40~km~s$^{-1}$ represent the thin
  disk population, and the $v_{\rm tan}$ = 150~km~s$^{-1}$ track
  illustrates the position of halo white dwarfs. The estimated
  empirical location of subdwarfs is also shown based on
  \citet{kilic10}. The regions labeled as A (dotted), B (short
  dashed), and C (long dashed) indicate the search criteria (see
  Table~1). 
  \label{fg:f_1}}
\end{flushright}
\end{figure}

\begin{figure}[p]
\epsscale{1.55}
\plottwo{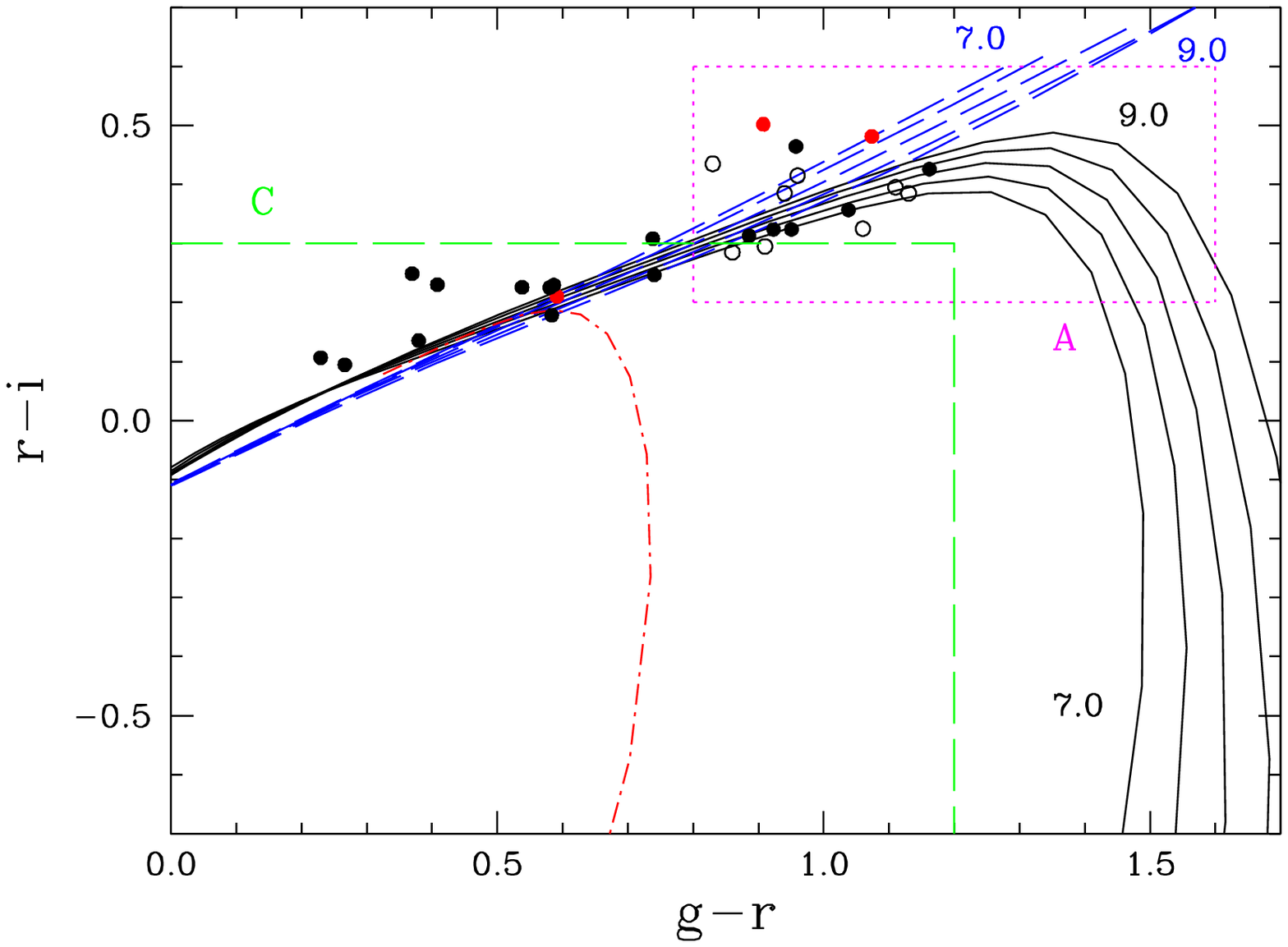}{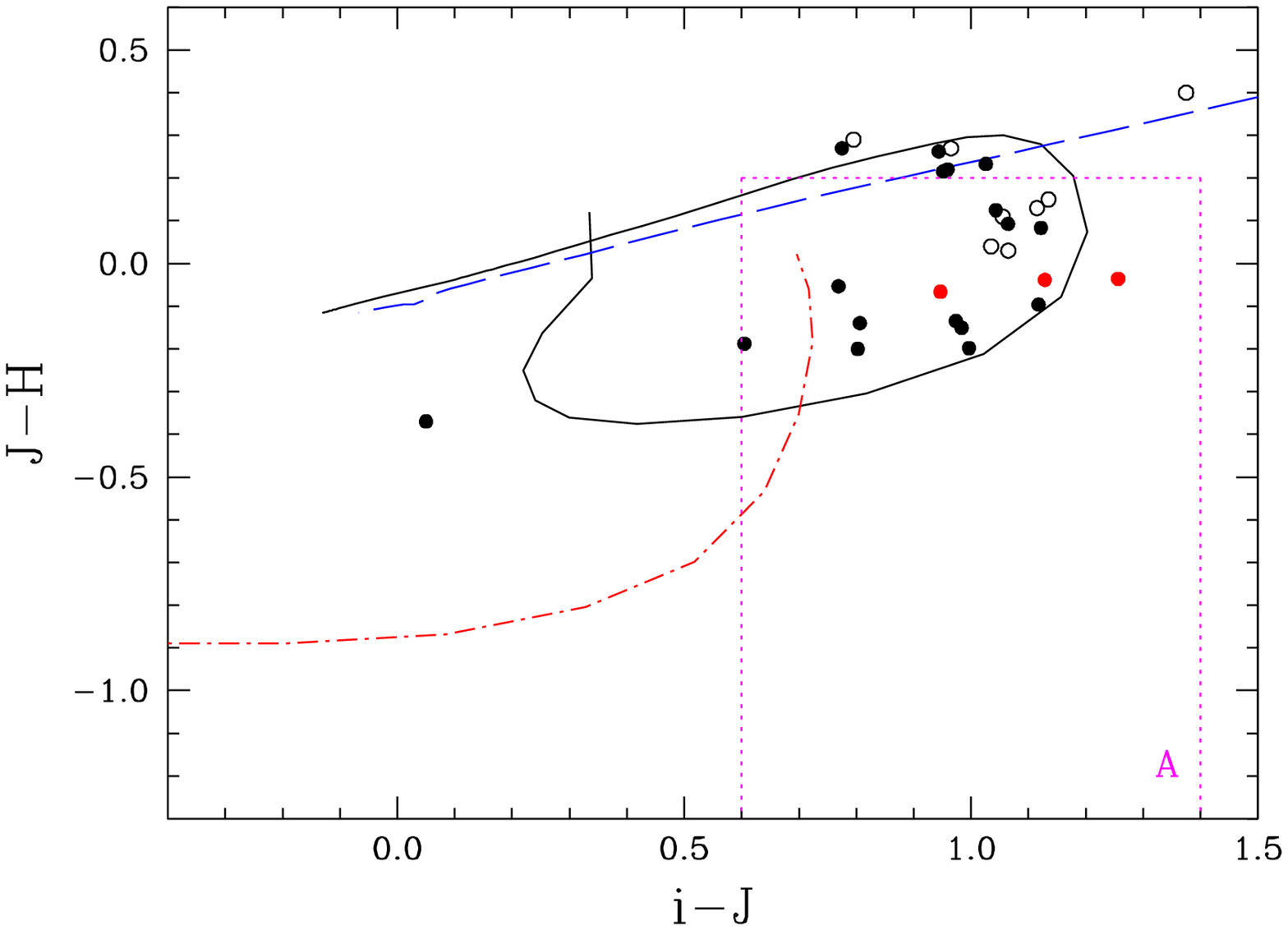}
\begin{flushright}
\caption{$r-i$ vs. $g-r$ (top) and $J-H$ vs. $i-J$ (bottom) color
  vs. color diagrams showing the color selection of white dwarf
  candidates. Model sequences at $\log g = 8$ are traced for pure-H
  atmospheres (solid, black), pure-He atmospheres (dashed, blue), and
  mixed He/H = 100 atmospheres (dot-dashed, red). On the top panel we
  also display pure-H and pure-He sequences from $\log g$ = 7.0 to 9.0
  with steps of 0.5~dex (extreme values are identified on the
  panel). We show the published sample of Paper~I and II (filled
  circles) and the new white dwarfs identified in this work (open
  circles). Mixed He/H objects are identified in red. The regions
  labeled A (dotted, magenta) and C (dashed, green) indicate the
  search criteria (see Table 1). The search region B covers the full
  range shown in the panels and search query C does not apply for
  objects with $H$ detection. \label{fg:f_2}}
\end{flushright}
\end{figure}

The color and reduced proper motion (RPM) selection of the published
LAS white dwarf candidates is discussed in Paper~I (Section 3) and II
(Section 2.3). In brief, the search was restricted to high
proper-motion sources with RPM values of $H_g = g + (5\log({\mu})+5) >
20.5$ (20.35 in Paper~I), where $g$ is the SDSS magnitude and ${\mu}$
the proper motion. The near-infrared flux was limited to $14 \le J \le
19.6$ to ensure non-saturation and a good detection while the
restriction $r_{\rm AB} < 20.7$ was employed to allow for
spectroscopic observations of the candidates. In Figures \ref{fg:f_1}
and \ref{fg:f_2} we present the sample of Paper~I and II in the
$H_{g}$ vs. $g-i$ RPM diagram, and $r-i$ vs. $g-r$ and $J-H$ vs. $i-J$
color versus color diagrams, respectively.

In Paper~I, sources as faint as 19.7 in $H$ were selected, although it
was found that LAS uncertainties were significantly underestimated for
such large $H$ magnitudes. In particular, many of the objects selected
as probable very cool white dwarfs with $J-H < -0.1$ were in fact
warmer degenerates with hydrogen-line spectra (DA) that were scattered
into the color selection because $H$ was too faint. This led to a more
stringent color selection and the acquisition of repeat $JHK$
photometry, using the WFCAM on UKIRT and NIRI on Gemini North, for
most candidates in Paper~II. It was confirmed that for $H \gtrsim 18$,
LAS $H$-band magnitudes are too faint by about 2$\sigma$.

We updated the photometry presented in Paper~I and II by relying on
the improved data reduction from the Data Release 10 of both the SDSS
and LAS surveys. We employ this improved data set to derive updated
atmospheric parameters in Section~4.

\subsection{New White Dwarf Candidates}

In this work, we have employed the LAS DR9 combined with the SDSS DR9
\citep{sdssdr9} to identify 18 new cool white dwarf candidates. The LAS
DR9 covers 3176 deg$^2$, most of which is also covered in SDSS DR9
since UKIDSS was originally designed to provide a near-infrared
counterpart to the SDSS survey. This suggests that we can
significantly enhance the size of the sample compared to the earlier
studies. We refer to Paper~II for a review of the derivation of the
proper motions from the combined SDSS/LAS data.

Table~\ref{tb:astro} gives the astrometric information for the new
candidates, and Table~\ref{tb:photo} presents the observed photometry
and associated uncertainties. While the sample was identified and
selected from the SDSS and LAS DR9, the photometric and astrometric
data used for our analysis and presented in Tables~\ref{tb:astro} and
\ref{tb:photo} is drawn from the DR10 of both surveys. It was found
that the new SDSS/LAS astrometry implies much smaller proper motions
for four objects, which led us to discard J0349$-$00 and J0838+24
since they are likely to be G- to K-type (sub)dwarfs ($H_g < 20$).

Table~\ref{tb:astro} identifies the three different RPM and color
regions where white dwarf candidates were selected in this
work. Region A corresponds to cool $T_{\rm eff} \sim 4000$~K pure-H
atmospheres, and this is where most white dwarfs were detected in
earlier studies. Out of ten candidates in this area, eight were
confirmed spectrocopically as white dwarfs (see Section 2.3), one is a
likely white dwarf with no spectrum, and one was found to be a G- to
K-type (sub)dwarf. Region B aims at recovering mixed H/He objects with
near-infrared CIA absorption, and we {\bf extend} this search area
compared to Paper II. This selection did not result in any new
confirmed white dwarf as all four candidates with $H_g > 20$ were
found to be either subdwarfs or extragalactic sources. We also created
a third search {\bf criterion} to identify objects with no $H$ or $K$
detection although it only resulted in one more object (J0916+30) as a
likely white dwarf, while one other was confirmed as a main-sequence
star. As also concluded in Paper II, the region A selection robustly
identifies white dwarfs with $T_{\rm eff}$ = 4100-4700~K, and cooling
ages of 7-9~Gyr.

{\bf Two of our candidates, J1240+25E and J1240+25W, have a close
  projected position on the sky. Over the period of 5.12 years covered
  between the SDSS DR10 and UKIDSS DR10 detections, the East component
  shows a proper motion of $+ 1\farcs 338$ in RA and $+ 0\farcs 103$
  in Dec. Over this same period the West component moves $+ 1\farcs
  348$ in RA and $+ 0\farcs 120$ in Dec. Given the close projected
  proximity, the movement in the same direction across the sky, and
  the agreement in the measured motion to within 4 mas yr$^{-1}$, we
  can identify the pair as a binary.}

\subsection{Additional Observations}

We obtained additional near-infrared photometry for a subset of the
sample where LAS data were missing. Repeat photometry was also secured
for objects where LAS uncertainties are larger than $\sim$ 0.1 mag
since the errors are likely to be underestimated in those cases (see
Paper II). The additional YJHK photometry was obtained for 12 targets
using NIRI \citep{hodapp03} on Gemini North. In Table \ref{tb:log} we
present the observation log and the data are given in Table
\ref{tb:niri}.

Optical spectroscopic observations were obtained for 13 objects and
eight of them turned out to be genuine white dwarfs. The Gemini
Multi-Object Spectrographs \citep[GMOS;][]{hook04} at both Gemini
North and South were used for all new observations. The R400 grating
was employed with the GG455 blocking filter and the central wavelength
was 680 nm, with wavelength coverage of 460-890 nm. As in Paper~II,
the 0.''75 slit was used with 2 $\times$ 2 binning and the resulting
resolution is R~$\sim1280$ or 5.5~\AA.

\begin{figure}[p]
\epsscale{0.9}
\plotone{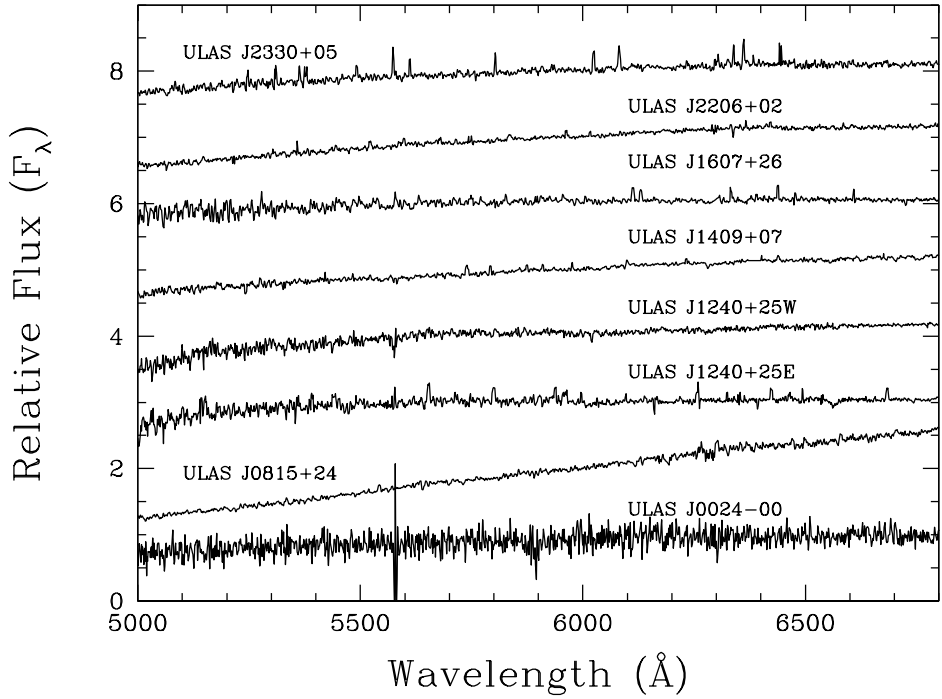}
\begin{flushright}
\caption{GMOS spectra of white dwarfs identified in the LAS DR9.  We
  have no spectroscopic observations for J0100$+$11 and J0916$+$30.
  The spectra have been normalized to unity at 6000~\AA~and offset in
  steps of one flux unit for clarity. \label{fg:f_spectra}}
\end{flushright}
\end{figure}

Figure \ref{fg:f_spectra} shows the GMOS spectra of the new white
dwarfs. All of them are featureless and consistent with the DC
classification except for J1240+25E which shows a weak
pressure-broadened H$\alpha$. In the following we add two objects
without spectra in our sample (J0100+11 and J0916+30) since the
photometric fits and tangential velocities suggest they could be white
dwarfs (see Section 4.2). We also rely on the white dwarf spectra
presented in Paper~I and II for objects found in earlier LAS data
releases.

The eight new spectroscopically confirmed white dwarfs and two
degenerate candidates are presented in the RPM and color vs. color
diagrams of Figures~\ref{fg:f_1} and~\ref{fg:f_2} along with white
dwarfs discovered in earlier LAS data releases. We also illustrate the
search regions A, B, and C. There is a significant number of new
objects outside of the search regions since we rely on the updated
DR10 samples and NIRI photometry instead of the DR9 used in the
initial selection. {\bf As in our earlier studies, some objects were
  scattered into the color selection because LAS $H$ was too faint.}

\section{MODEL ATMOSPHERES}

\subsection{1D Atmospheres}

We rely on pure-H, pure-He, and mixed composition 1D model atmospheres
that are similar to those described and used in Paper~I, Paper II,
\citet{kilic10}, \citet{giammichele12}, and
\citet{catalan12}. Contrary to earlier studies of LAS white dwarfs,
the pure-H models now include the Ly$\alpha$ red wing opacity from
\citet{kowalski06}, while the pure-He and mixed composition models are
identical to those described in \citet{kilic10}. The mixed composition
atmospheres have ratios of He to H by number ranging from values of
10$^{-2}$ to high values of 10$^{10}$.  All models rely on the
mixing-length theory (MLT; under the ML2/$\alpha$ = 0.8
parameterization) to treat convective energy transfer.

\subsection{Multi-Dimensional Effects}

For the coolest DA white dwarfs, there is no change in the predicted
spectra when the mixing-length parameterization of the MLT is varied
in the model atmospheres \citep{bergeron95}. The source of this
behavior is that the MLT predicts that convection is essentially
adiabatic throughout the atmosphere (i.e. there is no significant
entropy jump), in which case the thermodynamic structure only depends
on the value of the adiabatic gradient and not on the properties of
convective granules, such as the horizontal extent and lifetime.

It is desirable to confirm the robustness of the 1D approximation for
cool white dwarfs. For that purpose, we have computed CO$^5$BOLD
multi-dimensional simulations for $T_{\rm eff} = 3770$ and 4520~K, in
both cases at $\log g = 8$. We present the properties of these
simulations in Table~\ref{tb:3D}. The {\bf 2D} models use the same
numerical setup as the one described in \citet{tremblay13a} where the
coolest simulations are at $T_{\rm eff} = 6000$~K. {\bf On the other
  hand, the 3D simulations rely on a new version of CO$^5$BOLD with
  advances for low-Mach-number flows \citep{freytag13}. The improved
  numerical schemes are less diffusive and provide proper granulation
  in brown dwarfs (i.e. similar to models at higher Mach numbers)}. In
brief, we rely on an {\bf equation of state} and opacities that are
the same as those of our 1D model atmospheres described above. We
employ an 8 opacity bin scheme to describe the band-integrated
radiative transfer. The wavelength dependent opacities are sorted
based on the Rosseland optical depth ($\tau_{\rm R}$) at which
$\tau_\lambda$ = 1 and we take 150$\times$150$\times$150 grid points
to solve the radiative transfer and hydrodynamical equations. We have
derived $T_{\rm eff}$ from the temporally and spatially averaged
emergent stellar flux.

\begin{figure}[p]
\epsscale{0.9}
\plotone{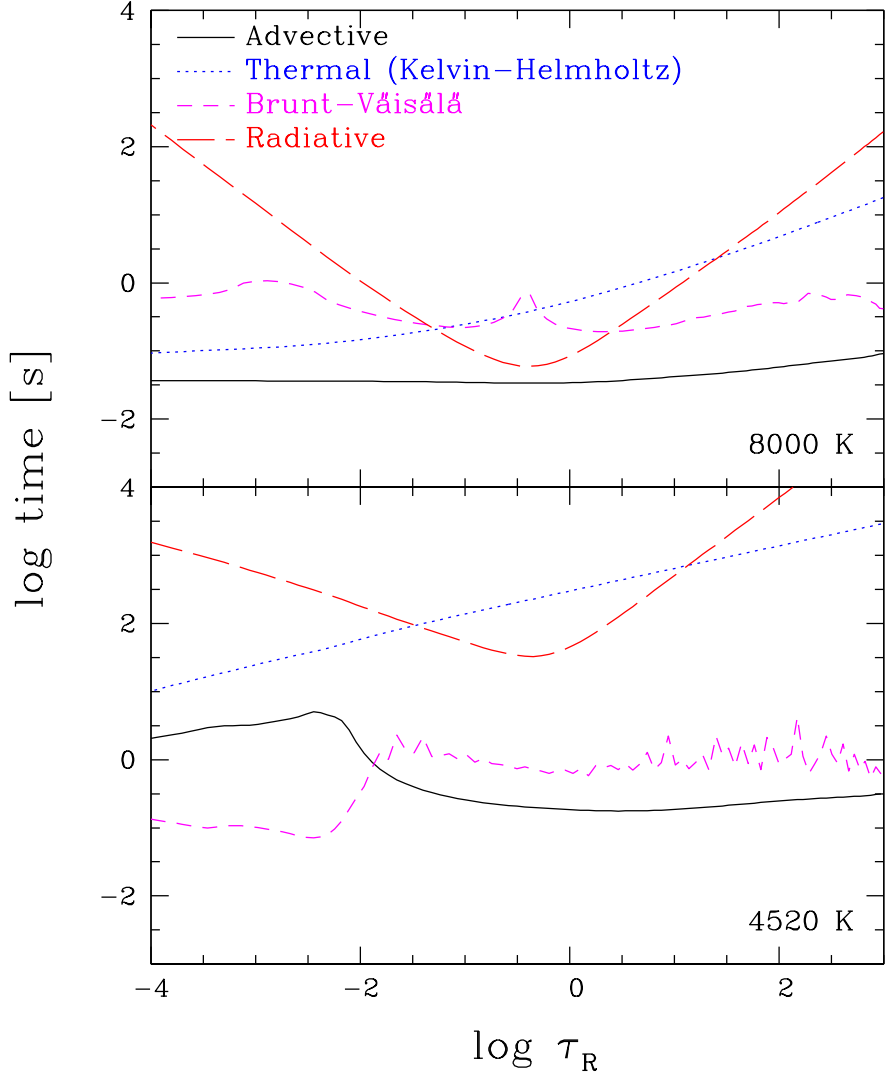}
\begin{flushright}
\caption{Timescales for the mean 3D structure of a 8000 (top panel)
  and 4520~K (bottom panel) pure-H white dwarf ($\log g = 8$) as a
  function of log $\tau_{\rm R}$. The different timescales are
  identified in the legend, and correspond to the advective timescale
  (black, solid), the Kevin-Helmholtz timescale (blue, dotted), the
  Brunt-V\"ais\"al\"a timescale (cyan, short dashed) and the radiative
  timescale (red, long dashed). See \citet[][Eq. 1-4]{tremblay13a} for
  the timescale definitions. Features at small optical depth ($\log
  \tau_{\rm R} < -2$) in the Brunt-V\"ais\"al\"a and advective
  timescales are likely numerical.
   \label{fg:f_time_ucool}}
\end{flushright}
\end{figure}

Figure~\ref{fg:f_time_ucool} demonstrates that the convective turnover
timescale is rather similar in very cool DC white dwarfs and hotter
DAs, with a value below one second. On the other hand, the radiative
relaxation timescale is more than 30 seconds in all layers at $T_{\rm
  eff} = 4520$~K which is rather long compared to DAs.  It implies
that radiative transfer will have little impact on the structure of
the convectively unstable layers. This decoupling between the two
timescales is also unfortunate for 3D simulations, since one must use
small time steps to solve the dynamical equations and have a long
total time to cover the radiative relaxation time. This problem is
similar to the one encountered in \citet[][see Section
  3.3]{tremblay13a} for cool DA white dwarfs ($T_{\rm eff} \lesssim $
8000~K). As in Tremblay et al. we perform instead non-gray 2D
simulations where it is possible to cover a few radiative relaxation
timescales. \citet{tremblay13a} have shown that most of the multi-D
effects are already in the 2D simulations, and that 3D mean structures
are only slightly different to their 2D counterparts.

\begin{figure}[p]
\epsscale{0.9}
\plotone{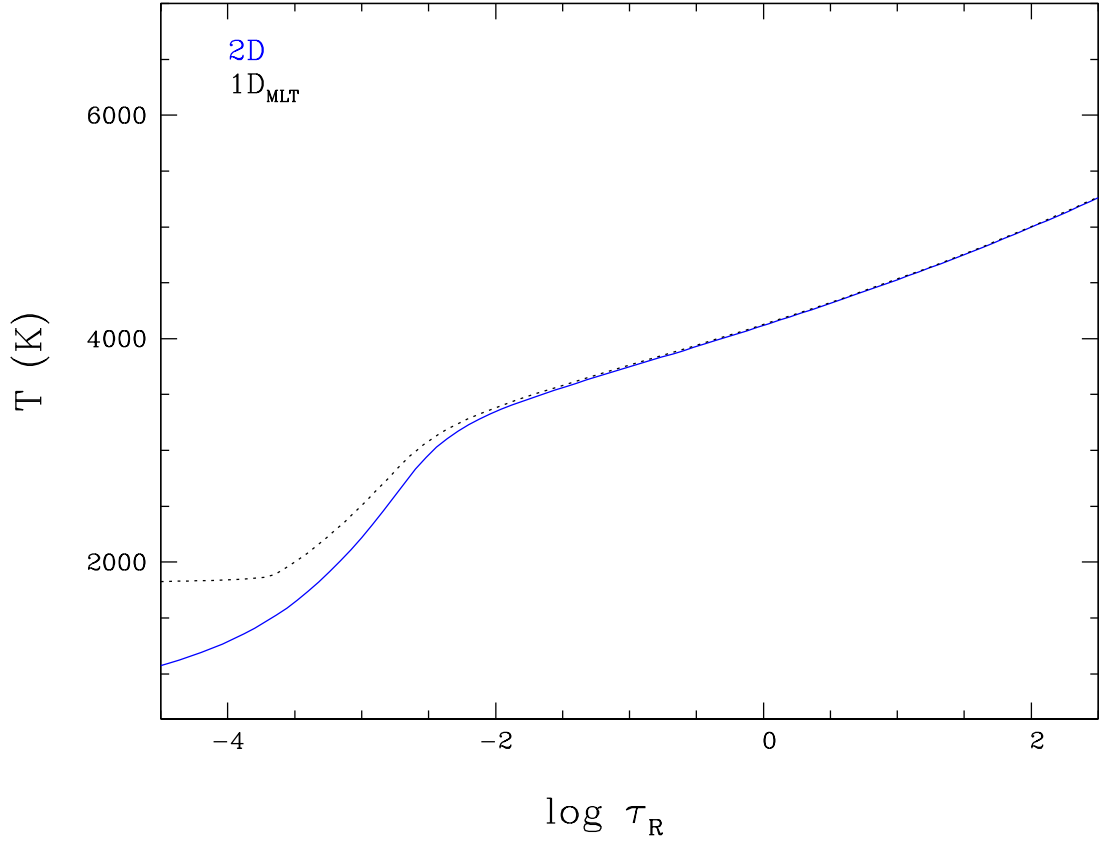}
\begin{flushright}
\caption{Non-gray temperature structure for a pure-H atmosphere at
  3770~K and $\log g$ = 8 from 2D (blue, solid) and 1D (black,
  dotted) models. \label{fg:f_adiabatic}}
\end{flushright}
\end{figure}

We find that both our models at $T_{\rm eff} = 3770$ and 4520~K
reach an almost completely adiabatic structure in all layers. This
outcome is similar to the coolest DA models in
\citet{tremblay13a}. The convective overshoot causes the entropy
gradient in the upper layers to relax to an adiabatic structure even
in regions that are stable to convection in 1D. In Figure
\ref{fg:f_adiabatic}, we compare 1D and $\langle$2D$\rangle$
structures for the simulation at 3770~K. The spatial and
temporal averages are performed over surfaces of constant Rosseland
optical depth and for 12 random snapshots. We observe
that both structures are nearly identical, except for the very top
layers. The results are very similar for the 4520~K case. In
these cool objects, H$_{2}$ molecule formation causes a strong
collision-induced absorption, and even the 1D model atmospheres are
unstable to convection up to $\tau_{\rm R} \sim 10^{-3}$. In the
convective layers, the structure is adiabatic and identical for 1D and
2D models since they rely on the same equation of state. Therefore, we
conclude that 1D model atmospheres relying on the MLT are appropriate
to model current observations of cool DC white dwarfs.

\subsection{3D Granulation Properties}

We have used our 2D simulations as initial conditions to compute 3D
atmospheres for 10 seconds in stellar time. Since radiative relaxation
timescales are much longer, we expect that the mean structure will
remain adiabatic and there is little interest in using the mean
properties of the 3D simulations. However, the 3D simulations allow
the study of the granulation properties, which is impossible in 1D and
rather limited in 2D.
 
\begin{figure}[p]
\epsscale{1.9}
\plottwo{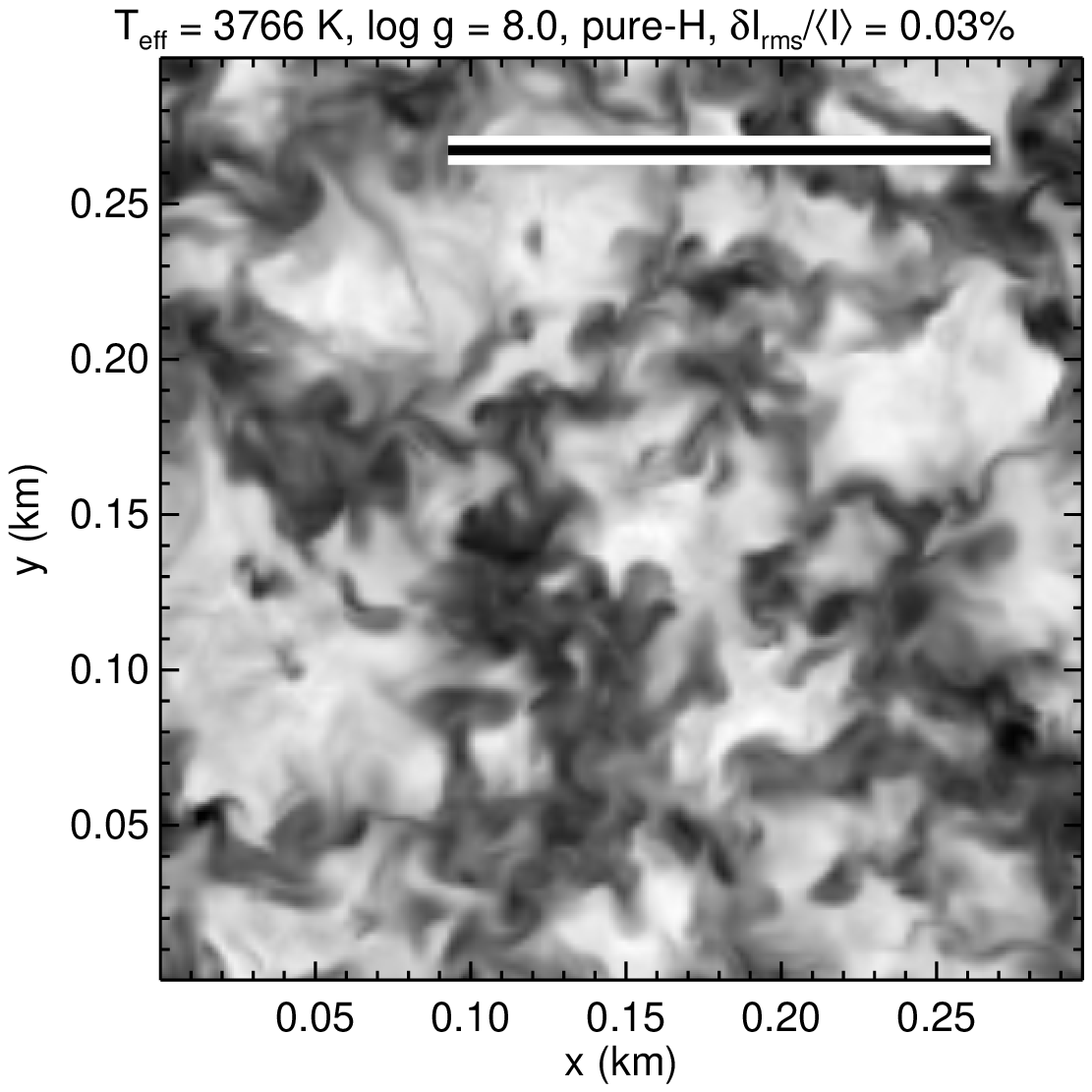}{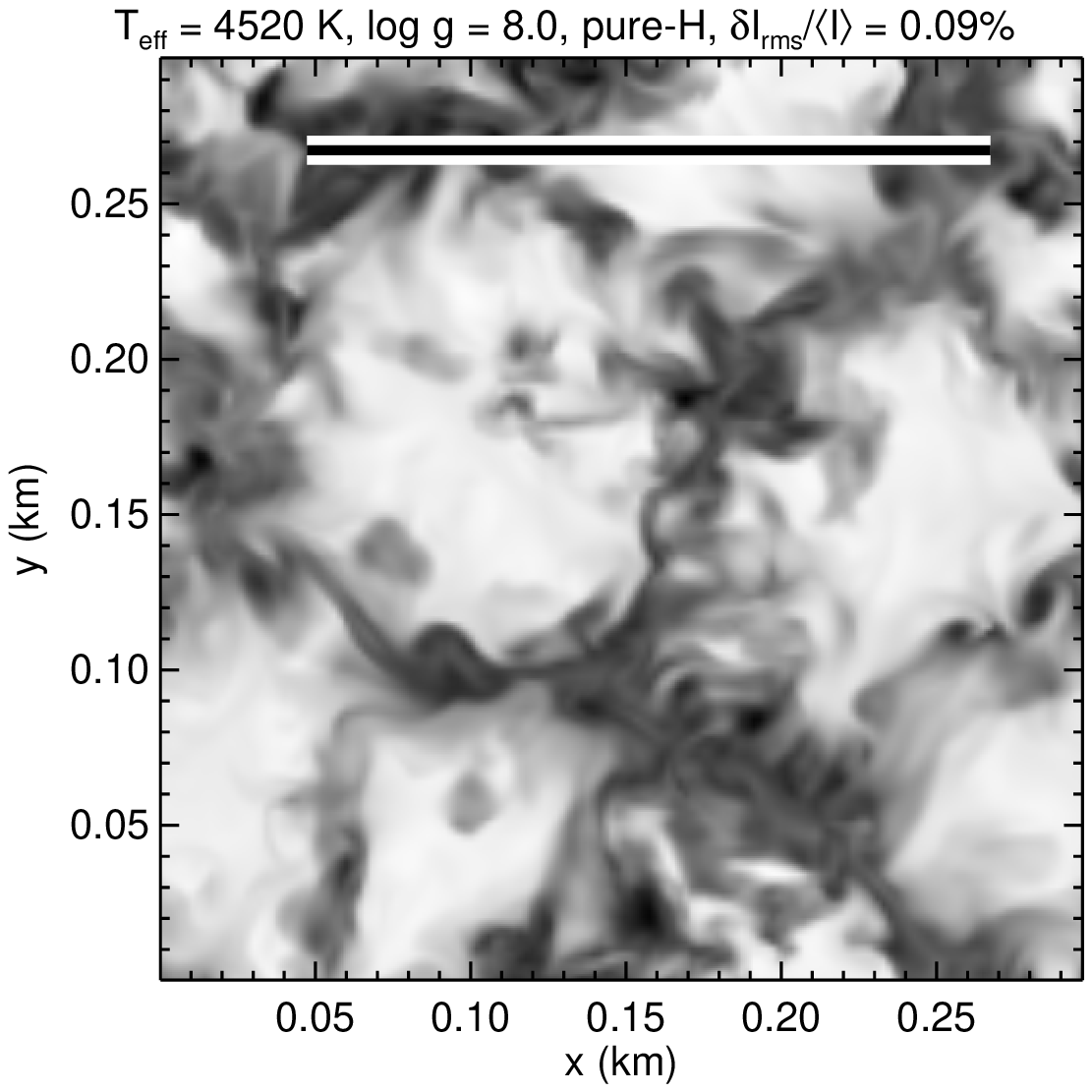}
\begin{flushright}
\caption{Emergent bolometric intensity at the top of the horizontal
  $xy$ plane of the 3D white dwarf simulations at $T_{\rm eff} =
  3766$~K ({\it top}), 4520~K ({\it bottom}), and $\log g = 8$. The
  root-mean-square intensity contrast with respect to the mean
  intensity ($\delta I_{\rm rms}/I$) is identified above the
  panels. The length of the bar in the top right is ten times the
  pressure scale height at $\tau_{\rm R} = 1$. \label{fg:f_3D}}
\end{flushright}
\end{figure}

Cool DC white dwarfs reach photospheric densities that are seen
otherwise in cool M dwarfs and brown dwarfs. However, the latter
objects contain metals in their atmospheres and are subject to
molecule, grain, and cloud formation in and above their photosphere
\citep{freytag10}. In Figure~\ref{fg:f_3D}, we present intensity
snapshots for our 3D simulations. The relative intensity contrast
\citep[see Eq.~73 of][]{freytag12} is given on the top of each
snapshot in Figure~\ref{fg:f_3D}. For our coolest simulation at
$T_{\rm eff} = 3770$~K, the intensity contrast is remarkably small at
0.03$\%$, although it is following the trend observed in
\citet{tremblay13b} given the low Mach number value in
Table~\ref{tb:3D}. In the same table we also give the characteristic
horizontal size of the granulation, which is roughly a factor of three
{\bf to four times} the pressure scale height at $\tau_{\rm R} = 1$,
in very good agreement with the trend found for cool DA white dwarfs
where the characteristic size reaches a plateau at around {\bf that
  value} for photospheric densities larger than $\sim 10^{-5}$ g
cm$^{-3}$.

\section{ANALYSIS AND RESULTS}

\subsection{Fitting Method}

In order to determine the atmospheric parameters of our observed
targets, we have first converted the magnitudes into observed fluxes
using the method of \citet{holberg06} and the appropriate filters. The
SDSS magnitudes are converted to the AB system employing the $u$, $i$,
and $z$ corrections given in \citet[][see their Section 2]{E06}. We fit the
resulting energy distributions with those derived from model spectra,
integrated over the same filters, using a nonlinear least-squares
method \citep{bergeron01}. Given our result that 3D effects are
negligible, in the following we rely only on the 1D spectra described
in Section 3.

From the fit of the observed energy distribution we constrain $T_{\rm
  eff}$ and the solid angle $\pi(R/D)^2$ where $R$ is the radius of
the star and $D$ is its distance from Earth. Since predicted energy
distributions depend only weakly on gravity and no parallax
measurements are available, it is not possible to constrain the
gravity of our objects. We therefore assume a surface gravity of $\log
g$ = 8. Most white dwarfs are found in a rather narrow range of
gravity \citep{bergeron92,kleinman13} and we employ, as in
Paper~II, a gravity uncertainty of $\pm 0.3$ dex. The photometric
variance uncertainties are obtained from the covariance matrix of the
fit.

We rely on the mass-radius relation from the evolutionary models of
\citet{fontaine01} with carbon-oxygen cores (50/50 by mass fraction
mixed uniformly) to determine the radius, mass, and cooling age. We
assume thick H layers ($M_{\rm H}/M_{\rm total} = 10^{-4}$) for
H-atmosphere white dwarfs and thin layers ($M_{\rm H}/M_{\rm total} =
10^{-10}$) for He and mixed atmospheres. We also derive the distance
to our targets by combining the mass-radius relation and the best-fit
solid angle. The $\log g = 8.0 \pm 0.3$ assumption dominates the
uncertainties on our derived masses and cooling ages. On the other
hand, it has little impact on the $T_{\rm eff}$ determinations for
which the uncertainty is dominated by the photometric variance.
Compared to Paper~II, we include the SDSS $u$ and $g$ filters for all
fits since our model spectra rely on the Ly$\alpha$ red wing
opacity. We use NIRI instead of LAS near-infrared photometry when
available, and neglect LAS photometry with uncertainties larger than
0.15 mag since it was found in earlier studies that errors were
significantly underestimated in those cases.

Finally, the surface composition is constrained from both the optical
spectra at H$\alpha$ and the photometry. Since H$\alpha$ absorption is
not seen in H-rich white dwarfs cooler than $\sim$5000~K, we can only
rely on the photometric fit in those cases. When possible, we also
estimate effective temperatures from a fit of the H$\alpha$ equivalent
width ($T_{\rm H\alpha}$).

\subsection{Stellar Parameters}

The best-fit atmospheric parameters for the ten new white dwarfs drawn
from the LAS DR9 and the 20 white dwarfs identified in Paper~I and II
are given in Table~\ref{tb:results}. We also derive distances and
cooling ages from evolutionary models as discussed in Section~4.1. The
distances and proper motions, given in Table~\ref{tb:astro}, are then
used to compute the tangential velocity $v_{\rm tan}$.

\begin{figure}[p]
\includegraphics[angle=270,scale=0.6]{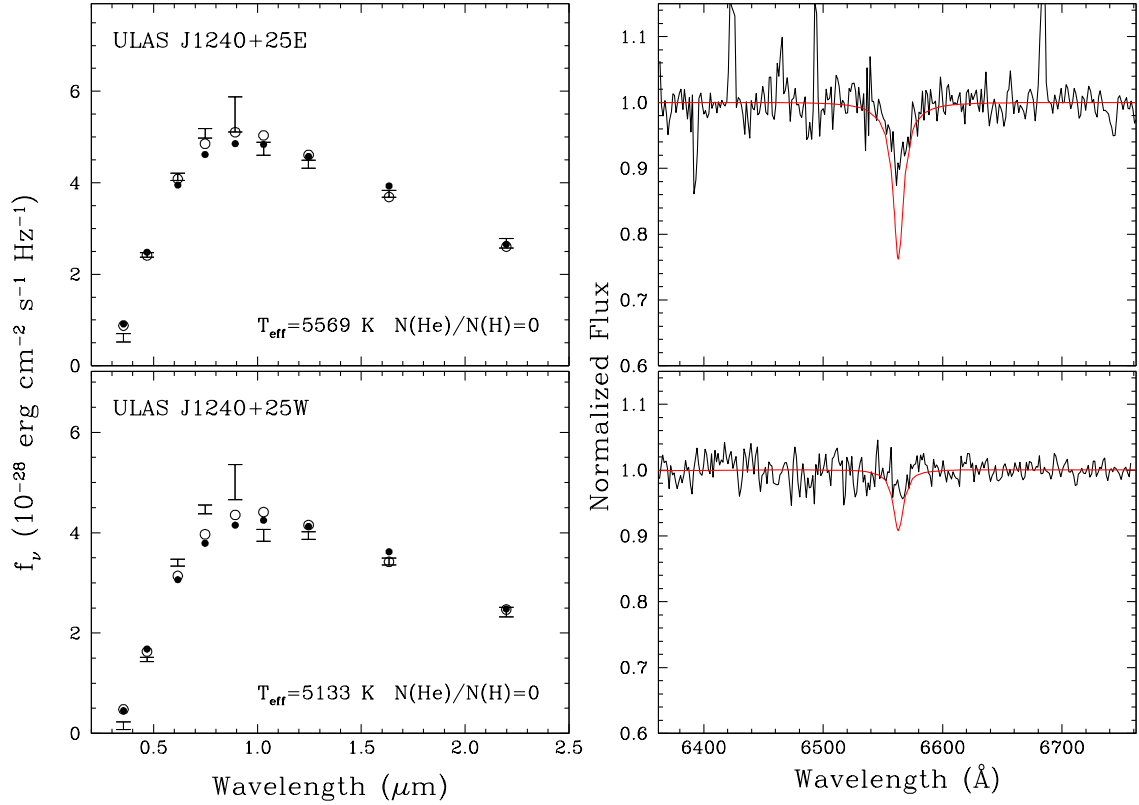}
\begin{flushright}
\caption{{\it Left panels:} Fit of the two components of a resolved
  white dwarf binary system from the LAS DR9 sample. The error bars in
  the left panels represent SDSS ugriz and LAS YJHK photometry.  The
  best fit models, averaged over the filter bandpasses, are shown for
  pure-H (filled circles) and pure-He compositions (open circles).  A
  surface gravity of $\log g$ = 8 is assumed, and the derived
  $T_{\rm eff}$ is shown in the legends for the selected
  composition. {\it Right panels:} Observed spectra around H$\alpha$,
  with the modeled pure-H atmosphere line profiles based on the
  atmospheric parameters derived from photometry (solid red line). The
  detection of H$\alpha$ in J1240+25W is
  uncertain. \label{fg:f_new_H_2p}}
\end{flushright}
\end{figure}

Among the newly identified white dwarfs, the fits shown in
Figure~\ref{fg:f_new_H_2p} for the resolved binary ULAS J1240+25E/W
are particularly interesting since we can compare the derived distance
and age of both components. J1240+25E is clearly H-rich given the
H$\alpha$ observation. However, while the photometric fit gives a
temperature of $\sim 5570$~K, the equivalent width of H$\alpha$ is
consistent with a $\sim$ 300~K lower $T_{\rm eff}$ value, which is
significantly larger than the variance allowed by the photometric
uncertainties. The photometric and $T_{\rm H\alpha}$ temperatures can
be {\bf brought into} agreement if the mass is very large
($>1~M_{\odot}$). The other component, J1240+25W, does not show a
clear H$\alpha$ absorption feature, and it is either a cooler H
atmosphere or a He-rich object. The fit quality is average for both
the pure-H (filled circles) and pure-He (open circles) cases, hence it
is difficult to constrain the composition only from the photometric
fit. However, the $u-g$ color has a large positive value, which is
consistent with a Ly$\alpha$ red wing absorption. We therefore select
the pure-H composition, also noticing that the spectrum shows a hint
of a faint H$\alpha$ line. The cooling age of the system is 3.0 and
5.0 Gyr based on the East and West component, respectively. We look at
this apparent discrepancy and compare the parameters of both
components of the system under different assumptions in Section
4.3. Nevertheless, it is clear that both white dwarfs have a
relatively low cooling age, and have a large velocity of $v_{\rm tan}
\sim 155$~km~s$^{-1}$, which could suggest disk remnants with a very
high velocity or former halo G stars.

\begin{figure}[p]
\epsscale{0.9}
\includegraphics[angle=270,scale=0.6]{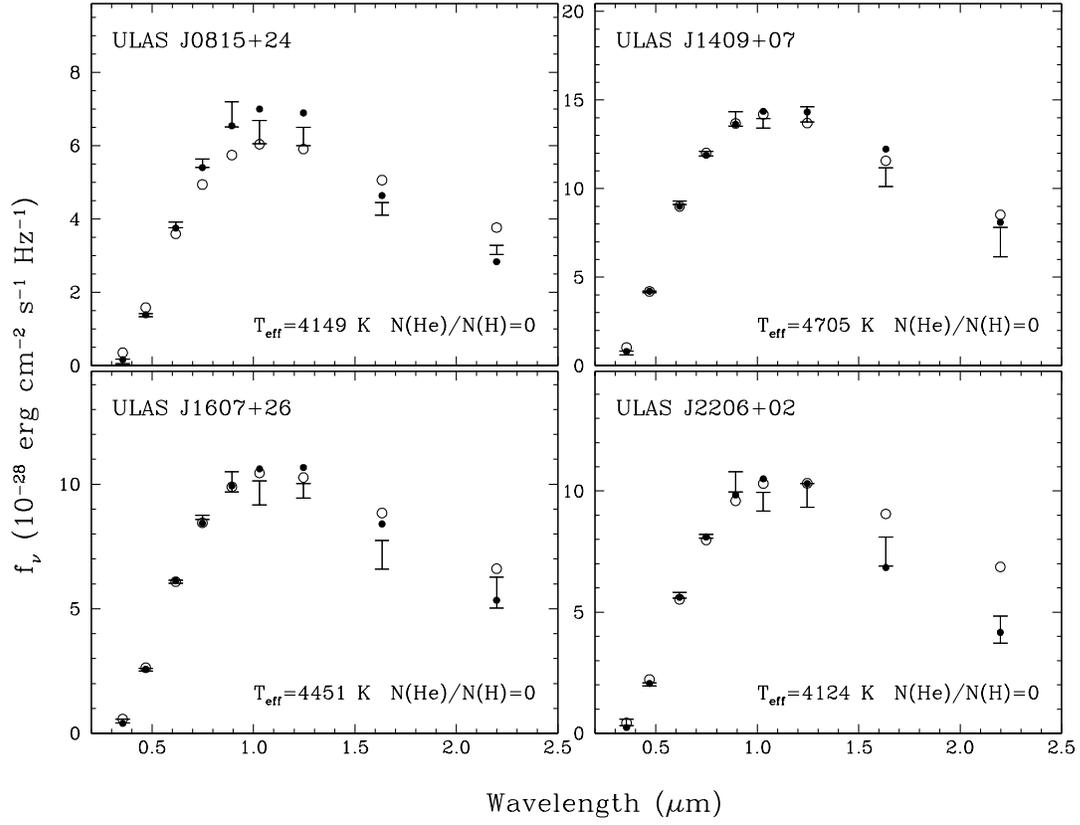}
\begin{flushright}
\caption{Four white dwarfs from the LAS DR9 sample best fit with
  pure-H atmospheres. Symbols are as in Figure~\ref{fg:f_new_H_2p}.
  All objects have observed and modeled featureless spectra (not
  shown).
  \label{fg:f_new_H_He}}
\end{flushright}
\end{figure}

\begin{figure}[p]
\includegraphics[angle=270,scale=0.6]{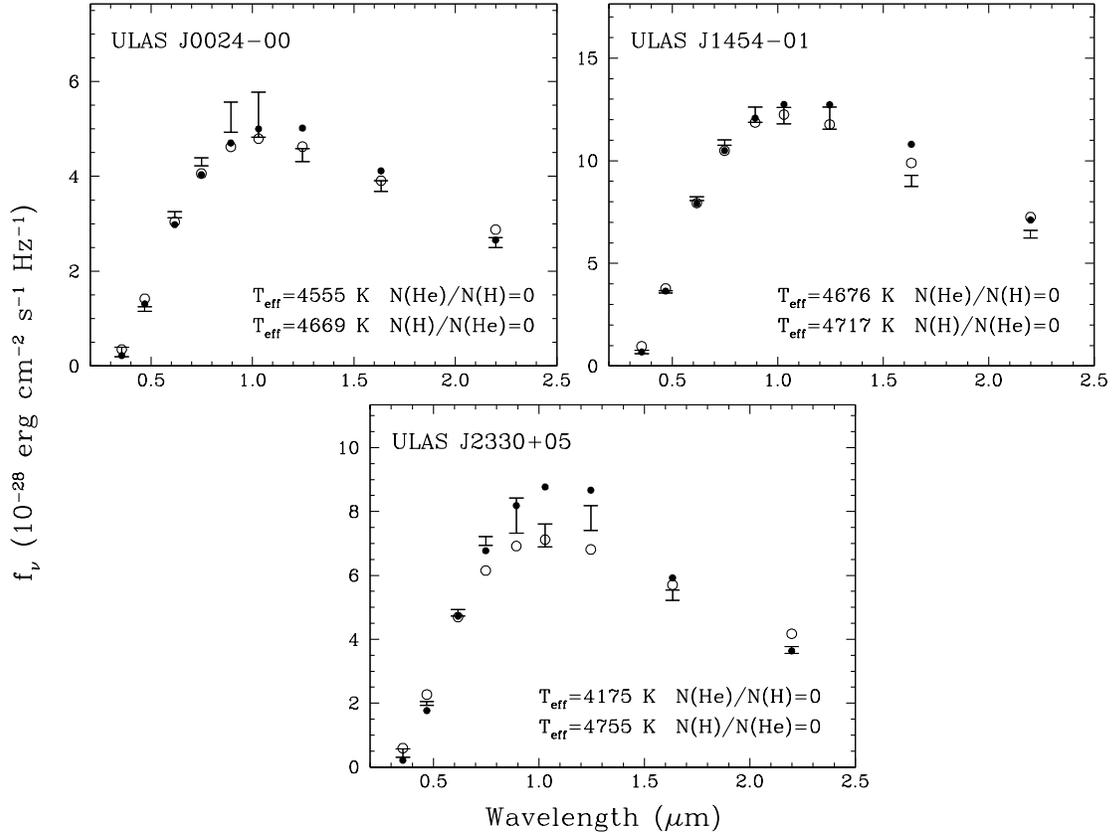}
\begin{flushright}
\caption{White dwarfs in the LAS DR9 sample and from
  \citet[][J1454$-$01]{paper2} with an unconstrained
  composition. Symbols are the same as in Figure~\ref{fg:f_new_H_2p}
  and we show the derived $T_{\rm eff}$ for both compositions in the
  legends.  The observed and modeled spectra are
  featureless. \label{fg:f_new_unc1}}
\end{flushright}
\end{figure}

\begin{figure}[p]
\includegraphics[angle=270,scale=0.6]{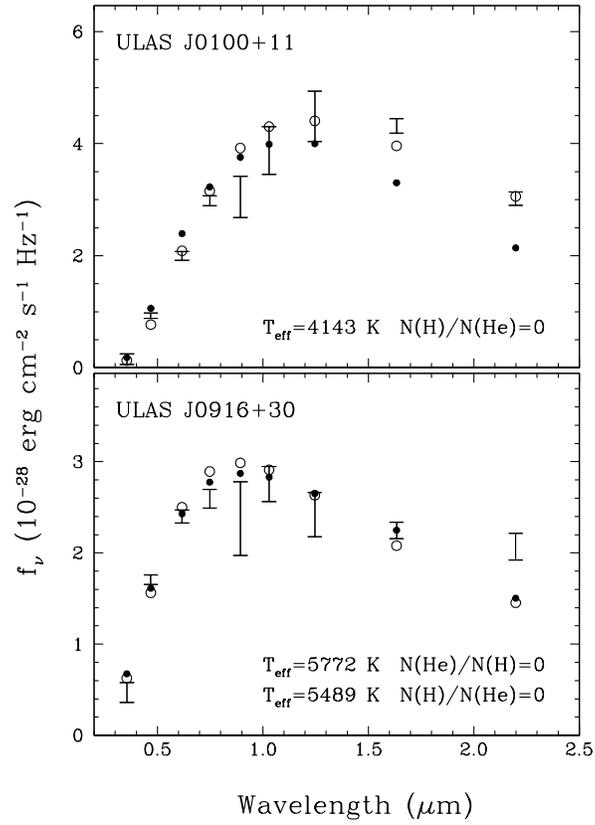}
\begin{flushright}
\caption{Two white dwarf candidates in the LAS DR9 sample with no
  spectroscopic observations. J0916+30 has an unconstrained
  composition.  Symbols are the same as in Figure~\ref{fg:f_new_unc1}.
\label{fg:f_new_unc2}}
\end{flushright}
\end{figure}

Figure~\ref{fg:f_new_H_He} displays four new white dwarfs that are
best fit with pure-H atmospheres. For J0815+24 and J2206+02, the
cooling ages of 8.5-9.0 Gyr and the tangential velocities in the range
40~km~s$^{-1}$ $\leq v_{\rm tan} \leq$ 60~km~s$^{-1}$ suggest they are
thick disk stars with a total age of 10-11 Gyr.
Figure~\ref{fg:f_new_unc1} exhibits two more new white dwarfs for
which we could not constrain the composition based on the photometric
fits and featureless DC spectra (J0024$-$00 and J2330+05).  Finally,
Figure~\ref{fg:f_new_unc2} shows two objects for which we do not have
spectra to confirm {\bf their} degenerate nature. J0100+11 has a
similar temperature to the coolest pure-hydrogen objects in our sample
in Figure~\ref{fg:f_new_H_He}, but a significantly larger
near-infrared flux that is better fitted with a pure-He model. This
would be one of {\bf the} coolest known white {\bf dwarfs} with a
derived pure-He composition if the nature of this object is confirmed.

\begin{figure}[p]
\includegraphics[angle=270,scale=0.6]{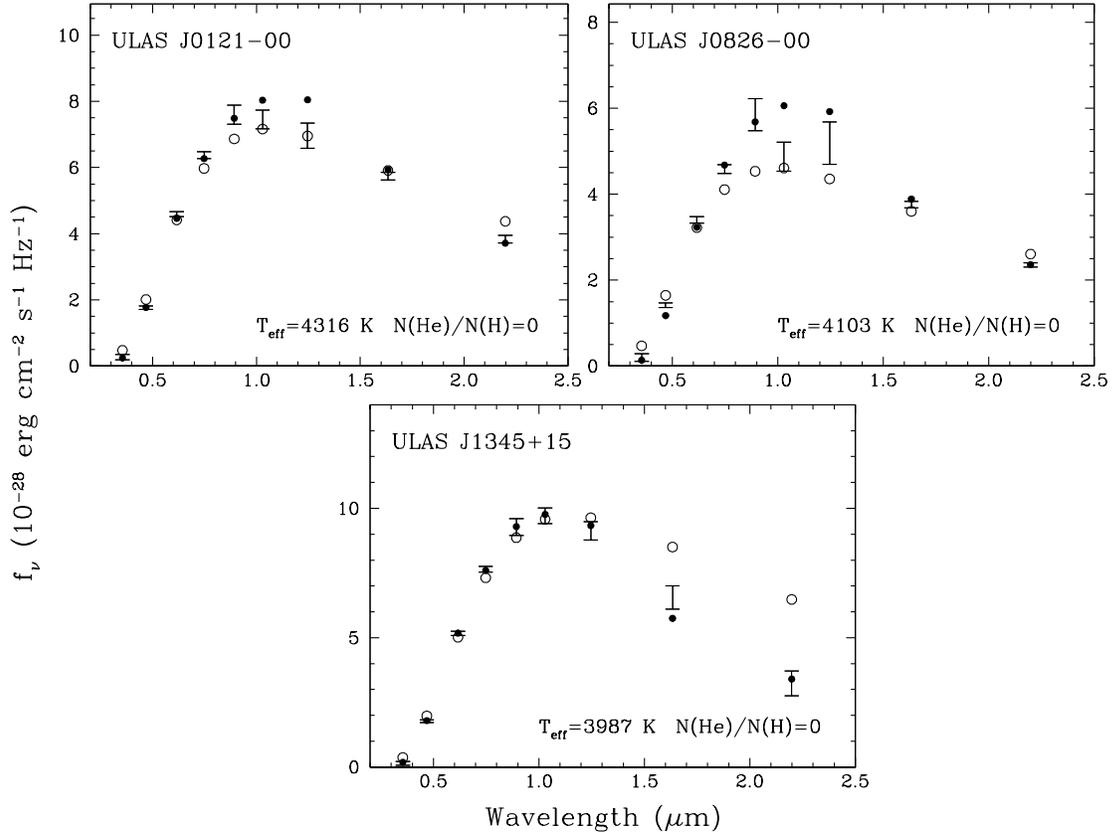}
\begin{flushright}
\caption{Improved fits for three objects from the \citet{paper2}
  sample with pure-H models (filled circles) including the Ly$\alpha$
  opacity \citep{kowalski06}. The derived $T_{\rm eff}$ is shown in
  the legends and we also display the pure-He composition best fit
  (open circles). \label{fg:f_old_H}}
\end{flushright}
\end{figure}

We have also performed a new analysis of the 20 LAS white dwarfs from
{\bf Papers}~I and II, with the main difference being that the pure-H models
now include the Ly$\alpha$ red wing absorption. Furthermore, we rely
on the more recently released SDSS and LAS DR10 photometry. The
derived atmospheric parameters are similar to those presented in
earlier works. This is in part because the $u$ and $g$ observations
were not included in the fitting procedure in earlier studies.  We
show three examples of cool pure-H degenerates in
Figure~\ref{fg:f_old_H} where the fit in the blue region is very good
for the pure-H solution, even though there is only a moderate change
in the atmospheric parameters compared to Paper II.

\begin{figure}[p]
\epsscale{0.9}
\plotone{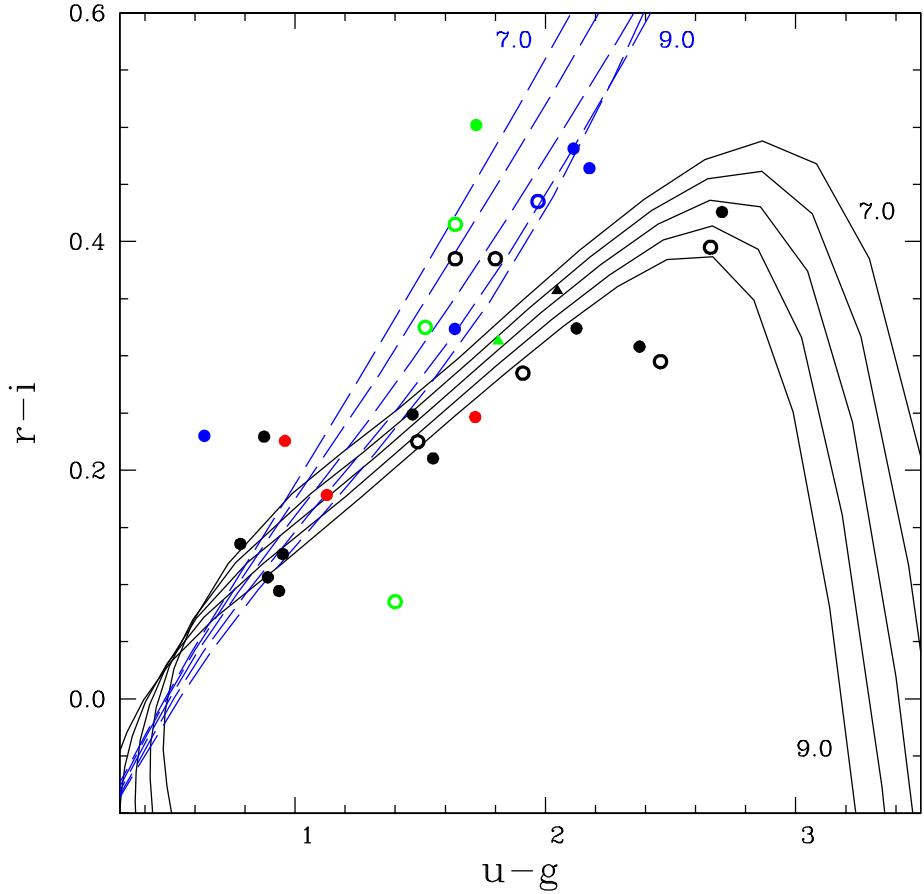}
\begin{flushright}
\caption{$r-i$ vs. $u-g$ color vs. color diagram for white dwarfs in
  the ULAS sample with a derived pure-H (black), pure-He (blue), mixed
  (red), or unconstrained (green) composition. The new white dwarfs
  identified in this work are shown with open circles while objects
  identified in Paper~I and II are identified by filled circles when
  the derived composition is unchanged, or with filled triangles in
  the two cases where we have updated the composition. Theoretical
  color sequences for pure-H (solid, black) and pure-He atmospheres
  (dashed, blue) are also shown for $\log g$ = 7.0, 7.5, 8.0, 8.5, and
  9.0 (extreme values are identified on the panel). \label{fg:f_umg}}
\end{flushright}
\end{figure}

The fits for pure-He and mixed objects are very similar to those of
Paper~I and Paper II. However, it was necessary to review the
composition assigned to the DC white dwarfs because of the improved
pure-H fits. Recent studies relying on the Ly$\alpha$ red wing opacity
have found that most DC white dwarfs below $T_{\rm eff} \sim 5000$~K
have a H-rich atmosphere \citep{kowalski06,kilic09,giammichele12}. In
Figure \ref{fg:f_umg}, we present the observed $r-i$ vs. $u-g$ color
vs. color diagram for our sample compared with theoretical sequences
for pure-H and pure-He atmospheres. The observed scatter is relatively
large, although this may be due to a scatter in the unknown surface
gravities. We remark that most objects identified as pure-He white
dwarfs in Paper~I and Paper II (blue filled circles) actually follow
the pure-He sequence (blue dashed lines) in Figure \ref{fg:f_umg} and
show no significant blue absorption. Hence, even by adding the blue
filters neglected in previous works, this does not impact the derived
composition for the three cool pure-He DC white dwarfs J1006+09,
J1206+03, and J1351+12.  However, we have updated the pure-He
classification for J1454$-$01, which is now unconstrained since the
pure-H fit is comparable to the pure-He fit, both shown in
Figure~\ref{fg:f_new_unc2}. Furthermore, J0121$-$00, which previously
had an unconstrained composition, is now better fit with a
pure-hydrogen atmosphere (displayed in Figure~\ref{fg:f_old_H} and
with a black triangle in Figure~\ref{fg:f_umg}).

We have derived $T_{\rm eff}$ values based on the equivalent width of
the H$\alpha$ line, when observed, that can be compared with the
photometric results. Since the strength of H$\alpha$ depends on the
unknown gravities, our values should be considered as estimates
only. For most of the DAs observed in Paper~I, $T_{\rm H\alpha}$ is
significantly lower than the photometric temperature. We note that
only $ugrizYJ$ was used in the photometric fit since the faint LAS $H$
for those objects was found to be unreliable. One explanation is that
the objects could be very massive white dwarfs, which would explain a
weaker H$\alpha$. Alternatively, it is possible that even the
uncertainties for the $ugrizYJ$ photometry have been
underestimated. We note that some of the objects with LAS photometry,
such as J0226$-$00, J1323+12, and J1522+08, reveal a good agreement
between photometric and H$\alpha$ temperatures.

\subsection{J1240+25E/W Binary System}

In this section, we compare the stellar parameters of the two cool
white dwarfs found in the resolved double degenerate system
J1240+25E/W. The objective is to better constrain the properties of
the system considering the fact that the total age and distance should
be the same for both objects. We have attempted to vary the gravity of
the components in order to match both the total ages and distances. To
compute the total age, we use the initial-final mass relation of
\citet[][]{kalirai08} with a third order polynomial fit, and the
main-sequence lifetimes of \citet{hurley00}. We reject solutions for
which the total age is larger than 13 Gyr. We only consider the
photometric variance in the uncertainties, hence we neglect the
effects of varying the initial-final mass relation. We therefore
consider that an agreement at the 3$\sigma$ level is acceptable.

\begin{figure}[p]
\epsscale{0.9}
\plotone{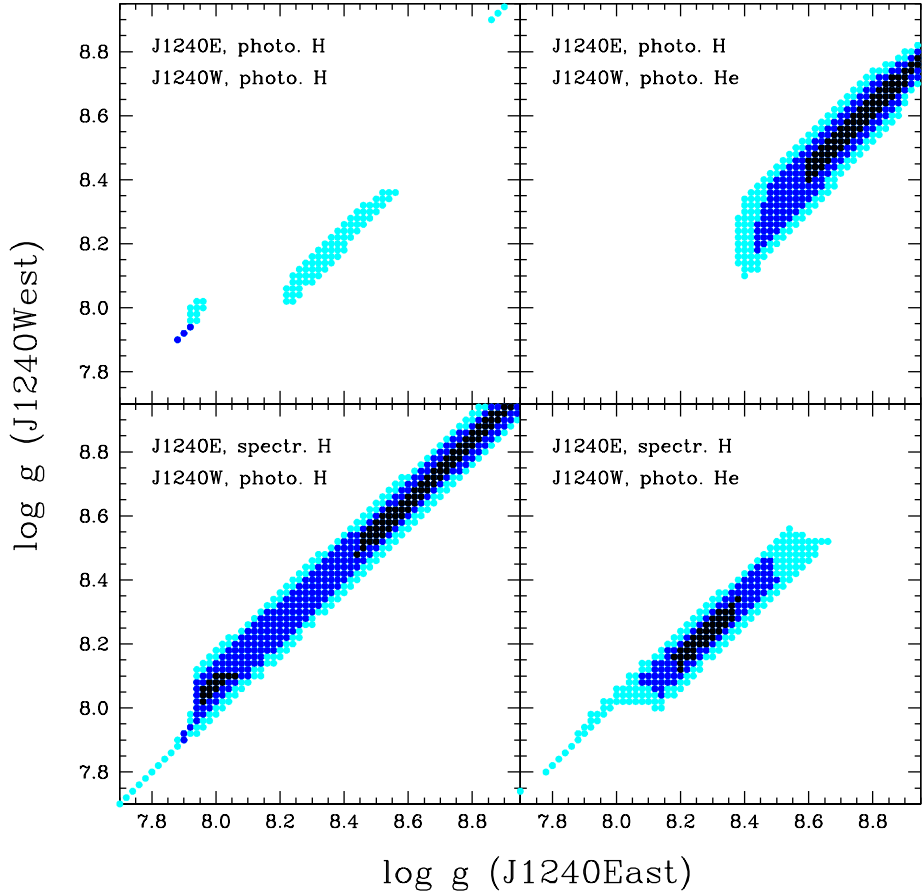}
\begin{flushright}
\caption{The $\log g$ couples for J1240+25E and J1240+25W where both
  the ages and distances are in agreement are shown when accounting
  for 1$\sigma$ (black circles), 2$\sigma$ (blue), and 3$\sigma$
  (cyan) of the photometric variance. The separate panels illustrate
  different assumptions about atmospheric parameters. We rely on the
  photometric and spectroscopic $T_{\rm eff}$ for J1240+25E on the top
  and bottom panels, respectively. J1240+25W is assumed to have pure-H
  and pure-He compositions on the left and right panels,
  respectively. \label{fg:f_binaryA}}
\end{flushright}
\end{figure}

In Figure \ref{fg:f_binaryA}, we display the $\log g$ couples for
which both total ages and distances are in agreement. Different
combinations are shown in terms of the uncertain $T_{\rm eff}$ value
of J1240+25E and composition of J1240+25W. The photometric $T_{\rm
  eff}$ solution for J1240+25E and pure-H composition for J1240+25W
(top-left panel) do not allow the ages and distances to both match
within 1$\sigma$ (black) although there are possible solutions within
2$\sigma$ (blue) or 3$\sigma$ (cyan). The bottom-left panel
demonstrates that when relying on the spectroscopic $T_{\rm eff}$ for
J1240+25E, the agreement is much better, and multiple solutions are
possible including the $\log g \sim 8$ canonical value for both
components. The pure-He solution for J1240+25W also provides matching
ages and distances, although only for higher gravities. Our results do
not provide strong constraints on the composition or atmospheric
parameters, but demonstrate that the apparent age discrepancy assuming
$\log g = 8$ can be resolved if we allow the gravities to
vary. Parallax measurements for the system will be necessary to
determine the distance and total age. The analysis also suggests that
the spectroscopic solution for J1240+25E is preferred, and that
underestimated photometric uncertainties are causing the discrepancy
between the spectroscopic and photometric $T_{\rm eff}$ solutions. We
notice from Figure~\ref{fg:f_new_H_2p} that the shape of the energy
distributions between $i$ and $Y$ are fairly different between the
observations and predictions. {\bf Since the problem is not generally
  observed in other data sets \citep[see, e.g.,][]{bergeron01}, it is
  likely an issue with the observations.}

\section{CHEMICAL EVOLUTION}
\subsection{Observational Status}

\begin{figure}[p]
\epsscale{0.9}
\plotone{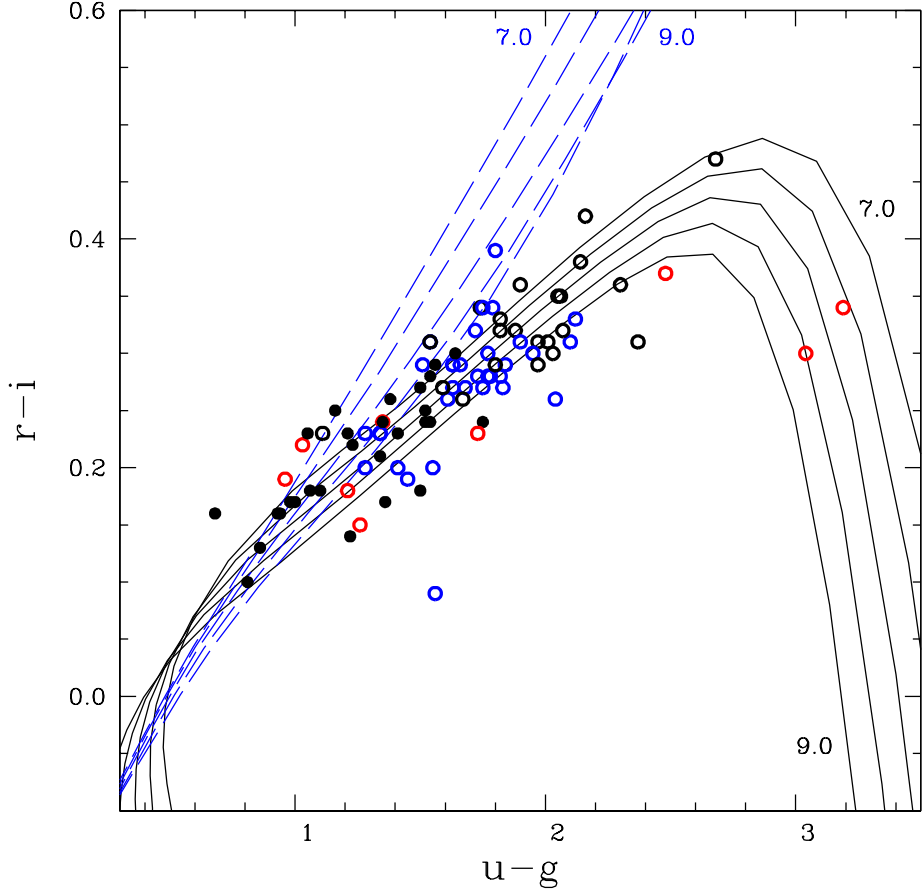}
\begin{flushright}
\caption{Similar to Figure~\ref{fg:f_umg} but for white dwarfs
  identified in the \citet{kilic10} sample. We only include objects
  with a DA (filled circles) or DC (open circles) spectroscopic
  classification. We also identify the objects for which Kilic et
  al. derived pure-H (black), pure-He (blue), and mixed (red)
  compositions. These classifications are with model atmospheres not
  including the Ly$\alpha$ red-wing opacity. \label{fg:f_umg_kilic}}
\end{flushright}
\end{figure}

In view of our results and those of similar recent studies, we review
the status of the spectral evolution of white dwarf atmospheres below
$T_{\rm eff} \sim 6000$~K. In order to have all studies on an equal
footing, we have refitted the $ugrizJHK$ photometry for the 126 cool
white dwarfs in \citet{kilic10} with models including the Ly$\alpha$
opacity. In Figure \ref{fg:f_umg_kilic}, we compare the observed $u-g$
colors for their sample with predictions from model atmospheres. Most
objects with a pure-He composition assigned in \citet{kilic10} are
actually in better agreement with the new pure-H sequence. However,
most of the pure-He objects are relatively warm ($T_{\rm eff} >
4600$~K) and in a regime where the predicted differences between the
pure-H and pure-He colors are close to the photometric uncertainties.
We have inspected the individual fits (not shown here), and the new
pure-H fits do not always provide an obvious better match to the
observations. We refrain from assigning new compositions to the
objects for the time being, although Figure~\ref{fg:f_umg_kilic}
demonstrates that it is a likely possibility that most derived pure-He
atmospheres with $T_{\rm eff} < 5000$~K in \citet{kilic10} are
actually H-rich atmospheres. In the following we consider both
possibilities.

In Table~\ref{tb:chemical}, we present the fraction of He-dominated
atmospheres ($N_{\rm He-dom}/N_{\rm total}$) found in this work, as
well as in the similar studies of \citet{giammichele12},
\citet{kilic09}, and \citet{kilic10}, respectively. We take the value
of He/H = 1 in number as the break point between the two compositions
in cases of mixed atmospheres. For the local sample of
\citet{giammichele12}, we only include objects within 20 pc, allowing
for a nearly volume complete sample and a better representation of the
local population.

The so-called non-DA gap in the range $5000 < T_{\rm eff} $ (K) $<
6000$ was first discussed in \citet{bergeron97}, where they found very
few pure-He atmosphere white dwarfs in that temperature range. In this
regime, it is fairly straightforward to define the main constituent of
the atmosphere, since the detection of H$\alpha$ and the agreement of
its observed equivalent width with predictions using the photometric
$T_{\rm eff}$ allows the confirmation of the pure-H interpretation. On
average, recent studies from Table~\ref{tb:chemical} suggest that only
$\sim25$\% of the white dwarfs in that $T_{\rm eff}$ range possess
He-dominated atmospheres. This confirms a deficit in He atmospheres
since for slightly warmer objects ($6000 < T_{\rm eff}$ (K) $< 8000$),
the ratio is in the range 40-50\% \citep{tremblay08,giammichele12}.

It was understood early that He-atmospheres are found in the non-DA
gap \citep[see the discussion in][]{bergeron01}, although in most
cases they are peculiar white dwarfs with pressure shifted C$_2$ bands
\citep[spectral type DQpec, see][]{kowalski10} or metals
(DZ). Furthermore, a significant fraction of cool He-dominated
atmospheres have a derived mixed He/H composition due to {\bf the}
presence of near-infrared CIA H$_2$-He absorption.  Indeed, for the
local white dwarf sample within 20 pc, \citet{giammichele12} find 25\%
of He-dominated atmospheres in the non-DA gap, and in the six
identified He-atmospheres, four are DQpec stars. The LAS sample is
smaller and does not allow for a statistically significant
characterization of the He-rich population in the non-DA gap, although
our results are in agreement with those of \citet{giammichele12}, with
a non-DA fraction of 25\% and one of the two He-rich objects is a
DQpec \citep[SDSS J1247+06;][]{kilic06}. The survey of \citet{kilic10}
is also in agreement with a non-DA gap, and more than half of the
He-dominated atmospheres (with spectroscopic observations) are DQpec,
DZs, or mixed He/H DCs with significant near-infrared
absorption. Finally, the sample of \citet{kilic09} is drawn from
\citet{bergeron01} and they did not include any of the DQpec {\bf
  stars}. Therefore, we believe that their low He-atmosphere fraction
is due to their target selection and we do not discuss further this
sample.

The non-DA gap described in \citet{bergeron01} has a red edge at
$T_{\rm eff} \sim 5000$~K. It is however difficult to differentiate
between pure-He and pure-H atmospheres at cooler temperatures since
both compositions have the DC spectral type. We exclude objects with a
derived $T_{\rm eff}$ below 4000~K since we lose the thin disk
contribution in this range, and most ultracool objects have peculiar
colors which are likely to lead to strong selection biases in
magnitude limited samples.

\citet{giammichele12} find that $N_{\rm He-dom}/N_{\rm total}$ = 24\%
in the range $4000 < T_{\rm eff}$ (K) $< 5000$ when employing the
Ly$\alpha$ opacity. All He-rich objects are DQpec, DZs, or mixed He/H
DCs, hence they find a white dwarf population that is very similar to
that found in the non-DA gap. This suggests that the non-DA gap is not
actually a gap but a one-time chemical evolution near $T_{\rm eff}
\sim$ 6000~K. The sample of \citet{kilic10} includes seven He-rich
mixed DC white dwarfs and one DZA in the range $4000 < T_{\rm eff}$
(K) $< 5000$, which implies a minimum ratio of $N_{\rm He-dom}/N_{\rm
  total} = 14 \%$ even if we consider that all derived pure-He
atmospheres are actually H-rich. Based on our new fits with models
including the Ly$\alpha$ opacity and the fact that the $u-g$ colors in
Figure~\ref{fg:f_umg_kilic} show little evidence of pure-He
compositions, it is unlikely that the number of He-dominated
atmospheres is much larger than the minimum fraction. Our sample has a
higher fraction of He-atmospheres in the same range, with a minimum of
18\% based on the fraction of mixed objects, but a value of 46\% if we
include the three DC white dwarfs that are better fitted with pure-He
models and if we exclude unconstrained compositions. This result is
certainly at {\bf odds} with the other samples, although our sample is
fairly small and the color selections favor the detection of mixed
objects. It is also possible that the discrepancy described in
Sections 4.2 and 4.3 between the observed photometry and predictions
has an impact on the derived compositions at low $T_{\rm eff}$. All in
all, there is no clear evidence with the most recent models and
observations, compared to the picture presented in \citet{bergeron01},
that the number of He-dominated atmospheres increases below $T_{\rm
  eff} = $ 5000~K.

\subsection{Theoretical Status}

In the following, we review the possible explanations for the spectral
evolution of He-atmospheres towards H-atmospheres at $T_{\rm eff} \sim
6000$ K. First of all, it could be caused by events in the interior,
perhaps in a way analogous to the convective mixing that is observed
from DA to DB between 30,000 and 20,000 K and from DA to DC at around
8000 K \citep{bergeron11,tremblay08}. The turnover timescales in white
dwarf convective zones, even in the deepest layers, are less than a
few hours \citep{grootel12}, hence many orders of magnitude shorter
than the characteristic cooling times. It appears improbable that
hydrogen could float on the surface due to incomplete mixing. However,
if the microscopic diffusion timescale of hydrogen in the helium-rich
plasma dominates over the turnover timescale somewhere within the
convective zone, it might be possible to create an abundance gradient.

According to \citet{tassoul90}, the convective zone in pure-He DC
white dwarfs stops growing at $M_{\rm He}/M_{\rm tot} \sim 10^{-6}$
once it reaches the degenerate core at around 10,000~K. In the highly
degenerate parts, conduction becomes the dominant energy transfer
mechanism, although convection is still predicted to take place in
layers with a relatively high level of degeneracy. Below $T_{\rm eff}
\sim 10,000$~K, the size of the convective zone slowly decreases due
to the growing degenerate core, although between 7000 and 5000~K, the
total mass included in the convective zone only decreases by $\sim$0.4
dex according to Table 4 of \citet{fontaine76}. This slow evolution is
in part because the level of degeneracy increases very slowly with
depth for adiabatic convection \citep[see Eq. 13 of][]{bohm68}. It is
therefore unlikely that changes in the composition of the atmosphere
would be linked to changes in the size of the convective zone. We note
that since the internal core is directly linked to the convective zone
in that regime, chemical evolution will lead to a change in the
relation between the surface and core temperatures. At a given initial
$T_{\rm eff}$, a He-atmosphere white dwarf turning to a H-atmosphere
would result in an object with a slightly lower $T_{\rm eff}$ due to
the larger difference between the core and atmospheric temperatures in
a DA white dwarf \citep{chen11}. This has to be taken into account in
a spectral evolution model.
 
\citet{chen12} put forward a scenario where a significant fraction of
H-atmospheres turn into He-dominated atmospheres at $T_{\rm eff} \sim
6000$~K, and the resulting DC objects are shifted to $\sim$500~K
higher $T_{\rm eff}$ by conservation of the core temperature. The
larger cooling rates for the post-mixing objects would create a
deficit of He-atmospheres below $6000$~K. In their scenario,
He-atmospheres would still be found in large number at $T_{\rm eff} <
5000$~K. We do not support this interpretation from the observational
evidences presented here. Furthermore, there is no strong {\bf
  constraint} on the total mass of hydrogen in DA white dwarfs to
suggest a DA to DC transformation around 6000~K and $M_{\rm H}/M_{\rm
  tot} \sim 10^{-7}$.  Instead we propose a DC to DA evolution from a
yet unknown mechanism.

Finally, the transition could be caused by an external factor, such as
the accretion of hydrogen from disrupted circumstellar material. Some
relatively warm ($T_{\rm eff} \sim$ 10,000~K) He-dominated white
dwarfs have already accreted a significant amount of hydrogen
\citep[see, e.g.,][]{zuckerman07}. From an extrapolation of the
accreted hydrogen masses from \citet{jura09} to cooler temperatures,
very cool DC white dwarfs with a history of accretion may have
hydrogen masses of the order of $M_{\rm H}/M_{\rm tot} \sim
10^{-8}-10^{-6}$. The upper limit gives a mass of hydrogen similar to
the mass of helium within the convective zone at $T_{\rm eff} \sim
6000$~K. While this could be the explanation for the presence of mixed
objects, this is insufficient to create a pure-hydrogen atmosphere,
and especially it would be very difficult to explain a rather abrupt
chemical evolution at $T_{\rm eff} \sim 6000$~K, which corresponds to a
cooling age of $\sim$2 Gyr.

The difficulties in defining a theoretical scenario for the chemical
evolution of very cool white dwarfs suggest that we may need to
further define the observed populations in larger samples and identify
the nature of the mixed, DQpec, and DZ degenerates as the first step.


\section{CONCLUSION}

The pairing of the LAS and SDSS samples, both from Data Release 9,
revealed eight to ten new white dwarf candidates from a
reduced-proper-motion and color selection. We have confirmed the
degenerate status of eight objects from spectroscopic observations
while two other objects without spectra could also be degenerate
stars. Our comparison of the observations with model atmospheres
suggests that most selected objects are very cool white dwarfs with
$T_{\rm eff} < 5000$~K. While we could not constrain the gravity from
the photometry alone, we have taken into account gravity effects in
our uncertainties. In three cases we could not constrain the
composition but we have confirmed two very cool pure-hydrogen remnants
with $T_{\rm eff} \sim 4100$~K and cooling ages between 8.5 and 9.0
Gyr. The tangential velocities in the range 40~km~s$^{-1}$ $\leq
v_{\rm tan} \leq$ 60~km~s$^{-1}$ suggest that they may be thick disk
objects. We have also identified the resolved binary ULAS J1240+25E/W
with two white dwarfs and an unusually large tangential velocity of
$\sim$ 155 km~s$^{-1}$. When relying on the spectroscopic H$\alpha$
temperature for J1240+25E, the total ages and distances agree for both
components although an unique solution can not be found without an
independent constraint on the gravities. The significantly hotter
photometric temperature for J1240+25E suggests that there are still
some unaccounted uncertainties in the SDSS and LAS photometric data at
the faint end.

We have reviewed the properties of the published LAS white dwarf
sample \citep{paper1,paper2} and updated two compositions given the
predictions of improved model atmospheres including the Ly$\alpha$ red
wing opacity. We have also computed 2D and 3D simulations but have
shown that the multi-dimensional effects are negligible for the
predicted spectra. The full LAS sample of 30 objects has been used to
study the spectral evolution from the number of helium- versus
hydrogen-dominated atmospheres. We found that our results are similar
to those of \citet{kilic10} and \citet{giammichele12} in the range
$5000 < T_{\rm eff} $ (K) $< 6000$ and we confirm a spectral evolution
towards pure-H atmospheres around $6000$~K. We have no theoretical
explanation for this chemical evolution. We argue that the transition
should be better defined from an observational point of view for a
sample with a well known completeness at faint magnitudes.

\acknowledgements

We are grateful for P. Kowalski for providing us the Ly$\alpha$
broadening profiles. Support for this work was provided by NASA
through Hubble Fellowship grant \#HF-51329.01 awarded by the Space
Telescope Science Institute, which is operated by the Association of
Universities for Research in Astronomy, Inc., for NASA, under contract
NAS 5-26555.

S.K.~L. is supported by Gemini Observatory. N.~L. was funded by the
Ram\'on y Cajal fellowship number 08-303-01-02\@. N.~L. is financially
supported by the project AYA2010-19136 from the Spanish Ministry of
Economy and Competitiveness (MINECO). This work was supported by
Sonderforschungsbereich SFB 881 "The Milky Way System" (Subproject A4)
of the German Research Foundation (DFG). This work is supported in
part by the NSERC Canada and by the Fund FRQ-NT (Qu\'ebec). J.~K. was
supported by the National Science Foundation (NSF) through grant
AST-1211719.

Some of the data reported here were obtained as part of the United
Kingdom Infrared Telescope (UKIRT) Service Programme; UKIRT is
operated by the Joint Astronomy Centre on behalf of the Science and
Technology Facilities Council of the U.K. This paper makes extensive
use of the UKIRT Infrared Deep Sky Survey (UKIDSS); we are grateful to
the WFCAM instrument team, the UKIRT staff, the UKIDSS team, the CASU
data processing team, and the WSA group at Edinburgh.

This paper also makes use of Sloan Digital Sky Survey (SDSS)
data. Funding for the SDSS and SDSS-II has been provided by the Alfred
P. Sloan Foundation, the Participating Institutions, the National
Science Foundation, the U.S. Department of Energy, the National
Aeronautics and Space Administration, the Japanese Monbukagakusho, the
Max Planck Society, and the Higher Education Funding Council for
England.

This work is based on observations obtained at the Gemini Observatory,
which is operated by the Association of Universities for Research in
Astronomy, Inc., under a cooperative agreement with the NSF on behalf
of the Gemini partnership: the National Science Foundation (United
States), the Science and Technology Facilities Council (United
Kingdom), the National Research Council (Canada), CONICYT (Chile), the
Australian Research Council (Australia), Minist\'erio da Ci\^encia e
Tecnologia (Brazil), and Ministerio de Ciencia, Tecnolog\'ia e
Innovaci\'on Productiva (Argentina).

\clearpage

 \clearpage
 \begin{deluxetable}{lrrrrrrr}
\tabletypesize{\footnotesize}
\tablewidth{0pt}
\tablecaption{Astrometry for Candidate White Dwarfs}
\tablehead{
\colhead{Short Name} & \colhead{Right Ascension} & \colhead{Declination} & \colhead{Epoch} & 
\multicolumn{2}{c}{$\mu$ $\arcsec$yr$^{-1}$} & \colhead{RPM} & \colhead{Search{\tablenotemark{a}}}\\
\colhead{} & \colhead{HH:MM:SS.SS} & \colhead{DD:MM:SS.S} & \colhead{Year} & \colhead{RA} & \colhead{Dec} &  \colhead{$H_g$} & \colhead{Region}\\
}
\startdata
ULAS J0024$-$00 {\tablenotemark{b}} & 00:24:39.96 & $-$00:30:39.6 & 2001.8808 & $-$0.051 & $-$0.160 & 22.33 & A \\
ULAS J0100$+$11 {\tablenotemark{c}} & 01:00:41.27 & 11:03:29.4 & 2008.8374 & $-$0.093 & 0.071 & 20.99 & A \\
SDSS J0113$-$02 {\tablenotemark{d,e}} & 01:13:55.25 & $-$02:57:29.9 & 2001.8616 & $-$0.05 & 0.39 & 24.64 & B \\
SDSS J0130$-$04 {\tablenotemark{d,e}} & 01:30:21.48 & $-$04:41:05.9 & 2008.9959 & 0.01 & 0.26 & 23.10 & B \\
ULAS J0156$-$00 {\tablenotemark{f}} & 01:56:09.78 & $-$00:14:50.4 & 2003.8863 & $-$0.107 & 0.044 & 22.25 & B \\
ULAS J0349$-$00 {\tablenotemark{c,g}} & 03:49:33.39 & $-$00:21:11.1 & 2003.7384 & $-$0.010 & 0.004 & 16.77 & B \\
ULAS J0815$+$24 {\tablenotemark{b}} & 08:15:53.48 & 24:27:35.0 & 2003.0781 & 0.107 & $-$0.059 & 21.49 & A \\
ULAS J0838$+$24 {\tablenotemark{c,g}} & 08:38:36.19 & 24:03:05.6  & 2003.9712 &  $-$0.015 & 0.040 & 18.17 & C \\
ULAS J0916$+$30 {\tablenotemark{c,g}} & 09:16:56.77 & 30:47:27.0 & 2004.1298 &  $-$0.046 & 0.057 & 20.14 & C \\
ULAS J1240$+$25W {\tablenotemark{b}} & 12:40:29.26 & 25:59:48.1 & 2005.0479 &  $-$0.263 &  $-$0.023 & 23.08 & A \\
ULAS J1240$+$25E {\tablenotemark{b}} & 12:40:29.48 & 25:59:47.6 & 2005.0479 &  $-$0.261 &   $-$0.020 & 22.54 & A \\
ULAS J1409$+$07 {\tablenotemark{b}} & 14:09:44.24 &  07:43:55.3 & 2003.3219 &  $-$0.215 &   $-$0.062 & 21.60 & A \\
ULAS J1607$+$26 {\tablenotemark{b}} & 16:07:38.95 & 26:08:47.2 & 2003.3247 & 0.046 & 0.156 & 21.44 & A \\
ULAS J1640$+$28 {\tablenotemark{f,g}} & 16:40:26.38 & 28:11:52.1 &  2003.4068 &  $-$0.036 & 0.039 & 19.59 & C \\
SDSS J2153$+$45 {\tablenotemark{d,f}} & 21:53:00.68 & 45:27:33.2 &  2006.4096 & $-$0.23 & 0.23 & 23.78 &  B\\
ULAS J2206$+$02 {\tablenotemark{b}} & 22:06:23.88 & 02:24:47.1 & 2008.7527 &  $-$0.172 & 0.123 & 22.27 & A \\
ULAS J2315$-$00 {\tablenotemark{f}} & 23:15:54.62 &  $-$00:13:32.8 & 2003.8863 & 0.025 & 0.069 & 20.93 & A \\
ULAS J2330$+$05 {\tablenotemark{b}} &23:30:22.55 & 05:40:22.2 & 2008.8156 & 0.136 &  $-$0.108 & 21.85 & A \\
\enddata
\tablenotetext{a}{Search queries are\\
A:  $H_g > 20.5$, $r < 20.7$, $ H < 18.9$, $0.8 \leq g-r \leq 1.6$, $0.2 \leq r-i \leq 0.6$, $0.6 \leq i-J \leq 1.4$, $J-H \leq 0.2$ \\
B: $H_g > 21.0$, $r < 21.0$, $J > 14.0$, $ r-i \leq [(0.5g-r)-0.4]$, $g-r \leq 2.0$, $r-i \leq 1.0$, $i-z \leq 0.5$, sdsstype$=$6\\
C: no $H$ or $K$ detection:  $H_g > 20.5$,  $r < 20.7$, $J \geq 19.0$, $y-j < 0.2$, $i-J < 1.0$, $r-i < 0.3$, $g-r < 1.2$}
\tablenotetext{b}{Confirmed as a white dwarf spectroscopically in this work.}
\tablenotetext{c}{No optical spectrum}
\tablenotetext{d}{Proper motion is determined from USNO catalog as specified in SDSS catalog}
\tablenotetext{e}{Confirmed as an extragalactic source spectroscopically in this work, implying incorrect proper motion.}
\tablenotetext{f}{Confirmed as a G- to K-type (sub)dwarf star spectroscopically in this work.}
\tablenotetext{g}{Earlier SDSS/UKIDSS astrometry implied larger proper motion.}
\tablecomments{The uncertainty in proper motion $\mu$ is $\sim$ 14
 mas yr$^{-1}$ (see Paper II). \label{tb:astro}}
\end{deluxetable}

%

\begin{deluxetable}{lrrrrrrrrr}
\tabletypesize{\scriptsize}
\tablewidth{0pt}
\rotate
\tablecaption{SDSS DR10 and UKIDSS LAS DR10 Photometry for Candidate White Dwarfs}
\tablehead{
\colhead{Short Name} & \colhead{$u$(err)} & \colhead{$g$(err)} & \colhead{$r$(err)} & \colhead{$i$(err)} &
\colhead{$z$(err)}  & \colhead{$Y$(err)} & \colhead{$J$(err)} & \colhead{$H$(err)} & \colhead{$K$(err)}  \\
}
\startdata
ULAS J0024$-$00 & 22.76(0.34) &  21.20(0.04) &  20.14(0.02)  & 19.80(0.02)  &  19.57(0.06)  &  18.96(0.09) &  18.71(0.13) &  18.56(0.21)   &   \nodata  \\    
ULAS J0100$+$11 & 23.49(0.63) &  21.48(0.05) &  20.65(0.04) &  20.20(0.03)  &  20.16(0.12)  &  19.30(0.11)  & 18.84(0.10)  & 18.77(0.25)    &  \nodata   \\
SDSS J0113$-$02 & 25.11(0.88) &  21.67(0.06) &  20.43(0.03) &  20.31(0.04)  &  20.57(0.16)  &     \nodata    &  \nodata  &   \nodata  &   \nodata  \\
SDSS J0130$-$04 & 22.97(0.45) &  21.02(0.04) &  19.91(0.02) &  19.79(0.02)  &  19.96(0.09)  &      \nodata   &   \nodata  &   \nodata  &   \nodata   \\
ULAS J0156$-$00 & 24.12(0.57) &  21.93(0.06) &  21.06(0.04) &  20.58(0.03)  &  20.19(0.10)  &  19.68(0.14) &  19.07(0.15)   &   \nodata  &   \nodata   \\
ULAS J0349$-$00 & 24.63(1.70) &  21.61(0.13) &  24.56(1.31)  &  24.22(1.33)  &  21.93(1.00)  &  19.80(0.12)  & 19.35(0.13) &  19.20(0.18)    &     \nodata   \\
ULAS J0815$+$24 & 23.75(0.54) &  21.05(0.03)  & 19.94(0.02) &  19.53(0.02)  &  19.28(0.05)  &  18.76(0.05) &   18.48(0.04)  & 18.63(0.21) &  18.37(0.27)   \\
ULAS J0838$+$24 & 21.00(0.07) &  20.02(0.02) &  19.88(0.02) &  19.85(0.03)  &  19.94(0.08)  &  19.32(0.05) &  19.16(0.08)  &    \nodata   &  \nodata   \\    
ULAS J0916$+$30 & 22.26(0.23) &  20.82(0.03) &  20.45(0.03) &  20.35(0.04)  &  20.43(0.17)  &  19.67(0.07) &  19.51(0.10)  &   \nodata   &  \nodata  \\
ULAS J1240$+$25W & 23.48(0.51) &  20.98(0.03) &  20.07(0.02) &  19.76(0.02)  &  19.62(0.07)  &  19.00(0.06) &  18.63(0.07) &  18.71(0.13) &  18.87(0.30) \\
ULAS J1240$+$25E & 21.97(0.15) &  20.44(0.02) &  19.86(0.02) &  19.62(0.02)  &  19.52(0.07)  &  18.80(0.05) &  18.67(0.07) &  18.35(0.09) &  18.33(0.18) \\
ULAS J1409$+$07 & 21.80(0.14) &  19.85(0.01)  & 18.99(0.01)  & 18.69(0.01)  &  18.51(0.03)  &  17.93(0.02) &  17.59(0.03) &  17.46(0.05) &  17.40(0.12) \\
ULAS J1607$+$26 & 22.22(0.16) &  20.38(0.02)  & 19.44(0.01) &  19.04(0.01)  &  18.86(0.04)  &  18.31(0.05) &  18.00(0.03) &  17.89(0.08)  & 17.63(0.11) \\
ULAS J1640$+$28 & 22.51(0.38) &  20.97(0.04)  & 20.62(0.04) &  20.45(0.05)  &  20.84(0.29)  &  19.82(0.13) &  19.73(0.22)   &  \nodata  &   \nodata    \\ 
SDSS J2153$+$45 & 22.93(0.53) &  21.22(0.05)  & 19.63(0.02)  & 19.28(0.02) & 20.56(0.19)      &    \nodata   &   \nodata  &   \nodata  &   \nodata   \\
ULAS J2206$+$02 & 22.32(0.30) &  20.64(0.03)  & 19.51(0.02) &  19.11(0.01)  &    18.83(0.04)  &  18.32(0.04) &  17.99(0.05)  &  17.84(0.08) &  17.93(0.13)   \\
ULAS J2315$-$00 & 22.89(0.27) &  21.60(0.05)  & 20.68(0.03)  & 20.17(0.03)  &  20.10(0.10)  &  19.31(0.08) &  18.98(0.08) &  18.87(0.18)  &   \nodata  \\
ULAS J2330$+$05 & 22.33(0.30) &  20.65(0.03)  & 19.69(0.02) &  19.26(0.02) & 19.13(0.07)   & 18.62(0.05)   &   \nodata    &  18.11(0.08)   & 18.24(0.18) \\
\enddata   
\tablecomments{SDSS $ugriz$ magnitudes are on the AB system \citep{AB}. LAS $YJHK$ are on the 
Mauna Kea Observatories (Vega) system \citep{vega}. \label{tb:photo}}
\end{deluxetable}

\clearpage

\begin{deluxetable}{lcccc}
\tabletypesize{\footnotesize}
\tablecaption{Observation Log}
\tablewidth{0pt}
\tablehead{
\colhead{Short Name} & \multicolumn{2}{c}{GMOS} &  \multicolumn{2}{c}{NIRI}\\
\colhead{} & \colhead{Date} & \colhead{Program} & \colhead{Date} & \colhead{Program}  \\
}
\startdata
ULAS J0024$-$00 {\tablenotemark{a}} &   &   & 2012-07-10  & GN-2012B-Q-109 \\
ULAS J0100$+$11 &   &   & 2012-07-20  & GN-2012B-Q-109 \\
SDSS J0113$-$02 &  2011-07-09  & GN-2011A-Q-98 &    &   \\
SDSS J0130$-$04  &  2011-07-26  & GN-2011A-Q-98 &    &   \\
ULAS J0156$-$00 & 2012-08-28  & GS-2012B-Q-72 &   2012-07-06 & GN-2012B-Q-109 \\
ULAS J0349$-$00 &   &   & 2012-10-09  & GN-2012B-Q-109 \\
ULAS J0815$+$24 & 2013-01-13,15  & GS-2012B-Q-72 & 2011-02-06  & GN-2011A-Q-59 \\
ULAS J0838$+$24 &   &   & 2011-01-25  & GN-2011A-Q-59 \\
ULAS J0916$+$30 &   &   & 2011-01-25  & GN-2011A-Q-59 \\
ULAS J1240$+$25W & 2012-12-19  & GN-2012B-Q-109 & 2013-01-16 & GN-2012B-Q-109 \\
ULAS J1240$+$25E & 2012-12-18 & GN-2012B-Q-109 & 2013-01-16 & GN-2012B-Q-109 \\
ULAS J1409$+$07 & 2011-05-13 & GS-2011A-Q-90 &  & \\
ULAS J1607$+$26 & 2011-05-01 & GN-2011A-Q-98 & & \\
ULAS J1640$+$28 & 2011-07-21 & GN-2011A-Q-98 & 2011-02-02 & GN-2011A-Q-59 \\
SDSS J2153$+$45 & 2011-07-04 & GN-2011A-Q-98 & &\\
ULAS J2206$+$02 & 2012-09-12,15 & GS-2012B-Q-72 &  &\\
ULAS J2315$-$00 & 2012-10-11 & GN-2012B-Q-109 & 2012-07-18  & GN-2012B-Q-109 \\
ULAS J2330$+$05 & 2012-07-26 & GS-2012B-Q-72 & 2012-07-10  & GN-2012B-Q-109 \\
\enddata
\tablenotetext{a}{Optical spectrum available in SDSS. \label{tb:log}}

\end{deluxetable}

\begin{deluxetable}{lrrrr}
\tabletypesize{\footnotesize}
\tablewidth{0pt}
\tablecaption{NIRI Photometry for LAS White Dwarf Candidates}
\tablehead{
\colhead{Short Name} &  \colhead{$Y$(err)} & \colhead{$J$(err)}  &  \colhead{$H$(err)} &    \colhead{$K$(err)}  \\
}
\startdata
ULAS J0024$-$00 & \nodata & 18.85(0.03) & 18.58(0.03) & 18.47(0.04) \\
ULAS J0100$+$11 & \nodata & \nodata & 18.44(0.03) & 18.31(0.04) \\
ULAS J0156$-$00 & 19.74(0.05) & 19.50(0.04) & 18.93(0.04) & 18.83(0.04) \\
ULAS J0349$-$00 & 19.80(0.05) & 19.65(0.04) & 19.05(0.04) & 18.99(0.05) \\
ULAS J0815$+$24  & \nodata & \nodata & 18.45(0.04) & 18.26(0.04) \\
ULAS J0838$+$24 & \nodata & \nodata & 19.15(0.03) & 19.18(0.03) \\
ULAS J0916$+$30 & \nodata & \nodata & 19.15(0.04) & 18.72(0.07) \\
ULAS J1240$+$25W & 19.28(0.03) & 18.98(0.02) & 18.69(0.02) & 18.55(0.04) \\
ULAS J1240$+$25E & 19.08(0.03) & 18.86(0.02) & 18.59(0.02) & 18.44(0.04) \\
ULAS J1640$+$28 &  \nodata & 19.69(0.04) & 19.41(0.07) &   \nodata \\
ULAS J2315$-$00 & \nodata & \nodata & 18.59(0.05) & 18.39(0.04) \\
ULAS J2330$+$05 & \nodata & \nodata & 18.20(0.03) & 18.10(0.03) \\
\enddata
\tablecomments{Photometry is on the Mauna Kea Observatories (Vega) system \citep{vega}. \label{tb:niri}}
\end{deluxetable}

 \clearpage
 \begin{deluxetable}{llllccccc}
 \tabletypesize{\scriptsize}
 \tablecolumns{9}
 \tablewidth{0pt}
 \tablecaption{CO5BOLD Simulations of Very Cool Pure-H DC White Dwarfs}
 \tablehead{
 \colhead{Dim.} &
 \colhead{$\Te$} &
 \colhead{$\logg$} &
 \colhead{time} &
 \colhead{$t_{\rm adv}$ ($\tau_{\rm R} = 1$)} &
 \colhead{$t_{\rm rad}$ ($\tau_{\rm R} = 1$)} &
 \colhead{$H_{\rm p}$ ($\tau_{\rm R} = 1$)} &
 \colhead{Char. size} &
 \colhead{Mach ($\tau_{\rm R} = 1$)}\\
 \colhead{} &
 \colhead{(K)} &
 \colhead{[cgs]} &
 \colhead{(s)} &
 \colhead{(s)} &
 \colhead{(s)} &
 \colhead{(m)} &
 \colhead{(m)} &
 \colhead{}
 }
 \startdata
2D & 3770 & 8.0 & 1000 & $-$    & $-$    &  $-$    & $-$    &  $-$    \\
3D & 3770 & 8.0 & 10  &  0.3  & 260 & 17 & 69  & 0.02 \\
2D & 4520 & 8.0 & 1000 & $-$    & $-$    &  $-$    & $-$    &  $-$    \\
3D & 4520 & 8.0 & 10  &  0.1  & 60  & 22 & 78  & 0.03 \\
 \enddata

 \tablecomments{All quantities were averaged over 250 snapshots and,
when appropriate, on the constant geometrical depth that corresponds
to $\langle \tau_{\rm R} \rangle_z = 1$. $T_{\rm eff}$ is derived from
the temporal and spatial average of the emergent flux, time is
the total simulation time, $t_{\rm adv}$ the advective or turnover
timescale, $t_{\rm rad}$ the radiative relaxation timescale, $H_{\rm
p}$ the pressure scale height, and Char. size the characteristic
horizontal size of the granulation. All quantities are defined in \citet{tremblay13b}. \label{tb:3D}}
\end{deluxetable}

 \clearpage
 \begin{deluxetable}{llllllll}
 \tabletypesize{\scriptsize}
 \tablecolumns{8}
 \tablewidth{0pt}
 \tablecaption{Derived Properties of the LAS Sample}
 \tablehead{
 \colhead{Short Name} &
 \colhead{Spectral} &
 \colhead{Composition} &
 \colhead{$T_{\rm eff}$} &
 \colhead{$T_{\rm H\alpha}$} &
 \colhead{Cool. Age} &
 \colhead{Distance} &
 \colhead{$v_{\rm tan}$}\\
 \colhead{} &
 \colhead{Type} &
 \colhead{} &
 \colhead{(K)} &
 \colhead{(K)} &
 \colhead{(Gyr)} &
 \colhead{(pc)} &
 \colhead{(km s$^{-1}$)}
 }
 \startdata
ULAS J0024$-$00   & DC    & Unconstrained & 4610 $\pm$ 130 & $-$  & 7.1$^{+2.3}_{-3.5}$ & 98 $\pm$ 21  & 78 $\pm$ 17\\
ULAS J0049$-$00   & DA    & H             & 6290 $\pm$ 90  & 5700 & 2.0$^{+1.8}_{-0.6}$ & 141 $\pm$ 26 & 83 $\pm$ 15\\
ULAS J0100+11     & $-$   & He            & 4140 $\pm$ 80  & $-$  & 8.0$^{+0.2}_{-3.0}$ & 88  $\pm$ 19 & 49 $\pm$ 11\\
ULAS J0121$-$00   & DC    & H             & 4320 $\pm$ 100 & $-$  & 8.2$^{+1.6}_{-2.8}$ & 71  $\pm$ 10 & 33 $\pm$ 5 \\
ULAS J0142+00     & DA    & H             & 6010 $\pm$ 100 & 5650 & 2.2$^{+2.0}_{-0.7}$ & 149 $\pm$ 27 & 74 $\pm$ 13\\
ULAS J0226$-$00   & DA    & H             & 5620 $\pm$ 120 & 5550 & 2.8$^{+2.4}_{-1.1}$ & 157 $\pm$ 28 & 82 $\pm$ 15\\
ULAS J0302+00     & DC    & He            & 5630 $\pm$ 150 & $-$  & 3.5$^{+2.2}_{-1.5}$ & 171 $\pm$ 34 & 99 $\pm$ 20\\
ULAS J0815+24     & DC    & H             & 4150 $\pm$ 100 & $-$  & 8.6$^{+1.4}_{-2.7}$ & 72 $\pm$ 11  & 42 $\pm$ 7 \\
ULAS J0826$-$00   & DC    & H             & 4100 $\pm$ 130 & $-$  & 8.8$^{+1.3}_{-2.6}$ & 76 $\pm$ 9   & 33 $\pm$ 4 \\
ULAS J0840+05     & DC    & Mixed         & 4360 $\pm$ 140 & $-$  & 7.5$^{+0.4}_{-3.0}$ & 82 $\pm$ 16  & 44 $\pm$ 9 \\
ULAS J0916+30     & $-$   & Unconstrained & 5630 $\pm$ 240 & $-$  & 3.3$^{+2.7}_{-1.6}$ & 168 $\pm$ 42 & 58 $\pm$ 15 \\
ULAS J1006+09     & DC    & He            & 4250 $\pm$ 110 & $-$  & 7.8$^{+0.4}_{-3.2}$ & 72 $\pm$ 17  & 50 $\pm$ 12\\
ULAS J1206+03     & DC    & He            & 4570 $\pm$ 70  & $-$  & 7.0$^{+0.7}_{-3.2}$ & 92 $\pm$ 19  & 72 $\pm$ 15\\
ULAS J1240+25E    & DA    & H             & 5570 $\pm$ 50  & 5250 & 3.0$^{+2.6}_{-1.2}$ & 130 $\pm$ 25 & 163 $\pm$ 31\\
ULAS J1240+25W    & DA?   & H             & 5130 $\pm$ 50  & $-$  & 5.0$^{+2.8}_{-2.5}$ & 124 $\pm$ 24 & 154 $\pm$ 30\\
ULAS J1320+08     & DC    & Mixed         & 4810 $\pm$ 130 & $-$  & 6.5$^{+0.8}_{-3.1}$ & 103 $\pm$ 19 & 96 $\pm$ 18\\
ULAS J1323+12     & DA    & H             & 5270 $\pm$ 70  & 5300 & 4.3$^{+2.8}_{-2.1}$ & 107 $\pm$ 19 & 106 $\pm$ 19\\
ULAS J1345+15     & DC    & H             & 3990 $\pm$ 80  & $-$  & 9.0$^{+1.4}_{-2.7}$ & 57 $\pm$ 9   & 71 $\pm$ 11\\
ULAS J1351+12     & DC    & He            & 4780 $\pm$ 60  & $-$  & 6.5$^{+0.9}_{-3.2}$ & 101 $\pm$ 20 & 58 $\pm$ 12\\
ULAS J1404+13     & DC    & Mixed         & 4350 $\pm$ 140 & $-$  & 7.5$^{+0.4}_{-2.7}$ & 59 $\pm$ 9   & 43 $\pm$ 7 \\
ULAS J1409+07     & DC    & H             & 4710 $\pm$ 30  & $-$  & 6.9$^{+2.1}_{-3.2}$ & 60 $\pm$ 11  & 64 $\pm$ 12 \\
ULAS J1436+05     & DC    & Unconstrained & 4320 $\pm$ 170 & $-$  & 7.9$^{+2.1}_{-3.7}$ & 56 $\pm$ 13  & 84 $\pm$ 19\\
ULAS J1454$-$01   & DC    & Unconstrained & 4700 $\pm$ 80  & $-$  & 6.8$^{+2.4}_{-3.4}$ & 63 $\pm$ 13  & 63 $\pm$ 13\\
ULAS J1522+08     & DA    & H             & 5370 $\pm$ 130 & 5450 & 3.7$^{+2.9}_{-1.7}$ & 151 $\pm$ 28 & 75 $\pm$ 14\\
ULAS J1607+26     & DC    & H             & 4450 $\pm$ 50  & $-$  & 7.7$^{+1.9}_{-3.3}$ & 64 $\pm$ 12  & 49 $\pm$ 9 \\
ULAS J2206+02     & DC    & H             & 4120 $\pm$ 70  & $-$  & 8.7$^{+1.6}_{-2.9}$ & 58 $\pm$ 10  & 58 $\pm$ 10\\
ULAS J2330+05     & DC    & Unconstrained & 4465 $\pm$ 380 & $-$  & 7.6$^{+2.4}_{-4.1}$ & 74 $\pm$ 23  & 61 $\pm$ 19\\
ULAS J2331$-$00   & DA    & H             & 6310 $\pm$ 130 & 5550 & 2.0$^{+1.8}_{-0.6}$ & 183 $\pm$ 34 & 119 $\pm$ 22\\
ULAS J2331+15     & DA    & H             & 5030 $\pm$ 60  & 5100 & 5.5$^{+2.7}_{-2.8}$ & 99 $\pm$ 20  & 75 $\pm$ 15\\
ULAS J2339$-$00   & DA    & H             & 6450 $\pm$ 200 & 5900 & 1.9$^{+1.7}_{-0.6}$ & 189 $\pm$ 35 & 92 $\pm$ 17\\
 \enddata

 \tablecomments{We assume $\log g = 8$ and uncertainties are computed
 from the photometric scatter and an allowed range in gravity of 7.7
 $\le \log g \le$ 8.3. The tangential velocity is calculated from the
 distance and proper motion given in Table~\ref{tb:astro} and Paper I
 and II. For J1323+12 we rely on the alternative proper motion given
 in Paper II. We do not include ULAS J1528+06 and J1554+08 from Paper
 I which were found to be subdwarfs. We also exclude LAS white dwarfs
 that were previously discovered in the SDSS (SDSS
 J2242+00; \citealt{kilic06}, and SDSS
 J1247+06; \citealt{kilic10}). \label{tb:results}}
\end{deluxetable}

 \clearpage
 \begin{deluxetable}{lcccc}
\tabletypesize{\footnotesize}
 \tablecolumns{5}
 \tablewidth{0pt}
 \tablecaption{Fraction of Helium Dominated Atmospheres}
 \tablehead{
 \colhead{$T_{\rm eff}$ range} &
 \colhead{LAS sample} &
 \colhead{Giammichele et al.} &
 \colhead{Kilic et al.} &
 \colhead{Kilic et al.} \\
 \colhead{(K)} &
 \colhead{(this work)} &
 \colhead{2012$^{a}$} &
 \colhead{2009} &
 \colhead{2010} 
 }
 \startdata
$7000 < T_{\rm eff} <8000$  & $-$     & 53\% & $-$ & $-$ \\
$6000 < T_{\rm eff} <7000$  & $-$     & 50\% & $-$ & $-$ \\
$5000 < T_{\rm eff} <6000$  & 25\%     & 25\% &  11\% & 30\%$^{b}$ \\
$4000 < T_{\rm eff} <5000$  & 46\%$^{c}$ & 24\% & 4\% & 14-59\%$^{d}$ \\
 \enddata
 \tablecomments{Helium dominated atmospheres are defined as DC, DZ, or DQpec with He/H~$>$~1. For completeness, we include the two SDSS white dwarfs re-discovered in the LAS sample (see notes in Table~\ref{tb:results}). \label{tb:chemical}}
\tablenotetext{a}{We only include objects within 20 pc.}
\tablenotetext{b}{We only include objects with a spectroscopic confirmation of the spectral type.}
\tablenotetext{c}{We do not include the two objects unconfirmed as white dwarfs as well as the unconstrained compositions.}
\tablenotetext{d}{The lower limit considers that all DC white dwarfs with derived pure-He atmospheres are actually pure-H atmospheres.}
\end{deluxetable}

\clearpage

\end{document}